%% file: main.tex
\documentclass[acmsmall,screen,nonacm]{acmart} %

\AtBeginDocument{%
  }

\setcopyright{acmlicensed}
\copyrightyear{2027}
\acmYear{2027}
\acmDOI{XXXXXXX.XXXXXXX}
\acmConference[FSE '27]{The 35th ACM International Conference on the
  Foundations of Software Engineering}{July 12--16, 2027}{Shenzhen, China}
\acmISBN{978-1-4503-XXXX-X/2027/07}

\input{mydefs}

\newcommand{\project}{\textbf{\textit{\textsc{ToFu}}}\xspace}

\begin{document}

\title{\project: Exploiting and Fuzzing the LLM--Orchestrator--Tool Trust
  Boundary}

\author{Md Nazmul Islam Ananto}
\affiliation{%
  \institution{Pennsylvania State University}
  \country{USA}
}
\author{Pierre Prevost-Avenel}
\affiliation{%
  \institution{Université Montpellier}
  \country{France}
}
\author{Colin Smith}
\affiliation{%
  \institution{Pennsylvania State University}
  \country{USA}
}
\author{Kyungtae Kim}
\affiliation{%
  \institution{Dartmouth College}
  \country{USA}
}
\author{Taegyu Kim}
\affiliation{%
  \institution{Pennsylvania State University}
  \country{USA}
}
\author{Arslan Khan}
\affiliation{%
  \institution{Pennsylvania State University}
  \country{USA}
}

\input{sec_abstract}




\maketitle

\input{sec_intro}

\input{sec_background}

\input{sec_moti}

\input{sec_design}

\input{sec_eval}

\input{sec_discussion}
\input{sec_conclusion}

\bibliographystyle{ACM-Reference-Format}
\bibliography{paper}

\clearpage
\appendix
\input{appendix}

\end{document}

%% file: mydefs.tex
\usepackage{amssymb}          %
\usepackage{mathtools}

\usepackage{wasysym}          %
\usepackage{upgreek}
\usepackage{pifont}

\usepackage{multirow}
\usepackage{makecell}
\usepackage{threeparttable}
\usepackage{rotating}      %
\usepackage{tablefootnote}

\usepackage{subcaption}       %
\usepackage{wrapfig}
\usepackage{tikz}
\usepackage{pgfplots}
\pgfplotsset{compat=1.18}

\usepackage{listings}
\usepackage{alltt}

\usepackage{algorithm}
\usepackage{algorithmicx}
\usepackage[noend]{algpseudocode}
\algrenewcommand\algorithmicindent{0.7em}

\usepackage{enumitem}
\usepackage{framed}
\usepackage{multicol}
\usepackage{xspace}
\usepackage{enumitem}

\colorlet{RED}{red}
\colorlet{BLUE}{blue}
\colorlet{ORANGE}{orange}
\colorlet{MAGENTA}{magenta}
\colorlet{CYAN}{cyan}
\colorlet{PURPLE}{purple}
\colorlet{OLIVE}{olive}

\definecolor{codekeycolor}{rgb}{0.11, 0.67, 0.84}
\definecolor{commentcolor}{rgb}{0.77, 0.12, 0.23}

\newif\ifdraftmarkers
\draftmarkersfalse

\newcommand{\cmark}{\CIRCLE}           %
\newcommand{\partialmark}{\LEFTcircle} %
\newcommand{\xmark}{\Circle}           %

\newcommand{\para}[1]{{\smallskip\noindent\textbf{#1~} $\,$}}

\lstdefinelanguage{C++}{
  morekeywords={uint8_t},
  sensitive=true,
  morecomment=[l]{//},
  morestring=[b]{"},
}

\makeatletter
\lst@InstallKeywords k{classattr}{classattrstyle}\slshape{classattrstyle}{}ld
\lst@InstallKeywords k{usertypeattr}{usertypeattrstyle}\slshape{usertypeattrstyle}{}ld
\makeatother

\usepackage{etoolbox}
\AtBeginEnvironment{table}{\renewcommand{\arraystretch}{0.9}}
\AtBeginEnvironment{table*}{\renewcommand{\arraystretch}{0.9}}

\usepackage{colortbl}
\definecolor{tofuteal}{named}{teal}
\newcommand{\tofurow}{\rowcolor{tofuteal!15}}

\definecolor{codebg}{gray}{0.92}
\definecolor{mathbg}{gray}{0.96}
\newcommand{\hlm}[1]{\begingroup\setlength{\fboxsep}{1pt}\colorbox{mathbg}{$#1$}\endgroup}
\usepackage{letltxmacro}   %
\LetLtxMacro{\origtexttt}{\texttt}
\newcommand{\ttsmall}[1]{\origtexttt{\fontsize{0.86em}{1em}\selectfont #1}}
\DeclareRobustCommand{\texttt}[1]{%
  \begingroup\setlength{\fboxsep}{1pt}%
  \colorbox{codebg}{\ttsmall{#1}}%
  \endgroup}
\AtBeginEnvironment{table}{\let\texttt\ttsmall}
\AtBeginEnvironment{table*}{\let\texttt\ttsmall}

%% file: sec_abstract.tex
\begin{abstract}
Large language models (LLMs) are increasingly deployed as agents that interact with external software through tools, with an orchestration layer translating model outputs into structured tool invocations.
This architecture creates an LLM--orchestrator--tool trust boundary where tool semantics may go unenforced and LLM-generated arguments may be incorrectly treated as trusted.
We present \project{}, a tool-aware framework that automatically derives semantic security checks for LLM tools from their existing metadata and a CWE-guided sanitizer corpus.
\project{} specializes these checks to individual tools and automatically generates fuzzing harnesses and semantic oracles using runtime introspection and outcome sanitization.
We evaluated \project{} across 273 MCP servers containing 3405 tools and identified verified bugs corresponding to CWE weakness classes in 343 tools across 118 of the servers.
Our findings show that real-world tools frequently fail to enforce their documented input semantics, making it unsafe to treat LLM-generated arguments as trusted.

\end{abstract}

%% file: sec_intro.tex
\section{Introduction}
\label{sec:introduction}
Large language models (LLMs) have demonstrated strong capabilities in
natural language understanding and generation. An \emph{orchestration
layer}~\cite{langgraph, langchain-tools, autogen} enables an LLM to act as an
\emph{agent} by connecting it to \emph{tools}, callable software functions
that interact with the external world. The layer supplies the LLM with
a user prompt alongside descriptions and input schemas of available
tools. The LLM uses this information to reason about the prompt, plan
actions, and generate structured tool calls specifying tool names and
arguments. The layer executes these calls and returns their results as
observations, allowing the LLM enabling agents to autonomously
perform complex tasks, including web browsing~\cite{openai2025operator},
software development~\cite{anthropic2025claudecode,wu2024devin},
and workflow automation~\cite{autogpt}.

While tools extend the capabilities of LLMs from generating text to
acting on external systems, their integration also exposes an attack
surface at the LLM--tool interface. Existing work has
investigated attacks that manipulate model behavior or tool use,
including jailbreaks~\cite{zou2023universal}, indirect prompt
injection~\cite{greshake2023not}, tool poisoning~\cite{mcp-scanner},
tool hijacking~\cite{toolhijacker}, and cross-tool
attacks~\cite{les-dissonances}.

Rather than focusing on prompt safety, we evaluate this boundary
from a systems perspective: \emph{What level of trust should tools
place in LLM-generated inputs?} Our first insight is that a tool
call is ultimately a structured model output that an orchestration
layer interprets as an executable request. We exploit this
interpretation through the \textit{Repeat-After-Me} attack.
Knowing the tool-call format recognized by the orchestration layer,
an attacker constructs text specifying a target tool and
attacker-chosen arguments, then instructs the LLM to repeat it
verbatim. Although the LLM merely reproduces the supplied text,
the orchestration layer interprets the resulting output as a tool
call and dispatches it for execution. A seemingly benign
text-repetition request thus gives the attacker direct control
over the tool invocation and its arguments.\footnote{We reproduced
this behavior across six evaluated orchestration layers:
LangGraph~\cite{langgraph},
MAF~\cite{microsoft-agent-framework}, ADK 2.0~\cite{google-adk2},
AutoGen~\cite{autogen}, Strands~\cite{strands}, and
CrewAI~\cite{crewai}.}
Conformity to a tool-call schema therefore does not establish
the trustworthiness of the resulting request. Due to lack of any authentication, \emph{LLM-generated tool arguments must remain
untrusted at execution time}. Rather than inducing unsafe model
outputs, we evaluate whether tool implementations safely handle
potentially malicious inputs.

Additionally, the metadata description provided by a tool and the actual implementation often does not align. Either a flashier promise is provided to the LLM so that this tool gets chosen or the tool hides malicious intent behind a benign description. In both cases, checking code implementation to establish both semantic integrity and trust in such cases of description-code inconsistency~\cite{shi2026description} is becoming more and more important.

Under this threat model, we seek to uncover vulnerabilities in tool
implementations that can be triggered by adversarial arguments.
Existing techniques, however, are poorly suited to this task at scale.
Current analyses of LLM tools primarily target predefined weakness
classes, specific taint flows, or particular agent--tool
attacks~\cite{mcp-first-glance,mcp-seclint,mcp-scanner,les-dissonances,viper-mcp,li2026don,guo2026systematic}.

Fuzzing provides a natural starting point for exploring a broader
range of adversarial inputs. However, conventional fuzzing oracles,
such as crashes, memory-safety violations, and assertion failures,
do not capture many tool vulnerabilities. A file tool may access
a resource outside its intended workspace, or a network tool may
contact a prohibited destination, while otherwise completing
successfully. Detecting these violations requires understanding
\emph{what an argument represents} and the security constraints
governing its use.

Our key insight is that \emph{the metadata that guides LLMs in using
tools can also guide fuzzers in testing them}. To support model-driven
invocation, tools expose descriptions of their functionality, argument
semantics, and intended usage. These descriptions provide semantic
context beyond input schemas: two arguments may both have type
\texttt{string}, yet one represents an SQL query and another a URL,
requiring different adversarial inputs and security checks. We
repurpose this metadata as a source of candidate security
specifications, using it to guide both adversarial input generation
and the construction of oracles that detect violations of the
corresponding security constraints.

We realize our insights by designing \project{}, a tool-aware fuzzing framework for automatically discovering semantic security vulnerabilities in LLM tools. To achieve this, \project{} introduces
\emph{semantics-aware sanitizers} (\S\ref{sec:design-sanitizer}), a reusable corpus that associates semantic
input classes (\texttt{path/URL/SQL}etc.) with CWE-guided security properties, adversarial inputs, and the runtime as well as outcome observations needed to identify such violations. Given a registered tool,
\project{} maps its arguments to applicable sanitizers 
using the semantics \& purpose documented in the tool metadata. \project{} then automatically generates schema-valid fuzzing harnesses and
semantic oracles. The use of an LLM, in \project{}, is confined within - assistance with interpreting natural-language metadata to map the sanitizers and generate initial fuzzing seeds only. The LLM is not used as the security oracle or judge. When implementation-level visibility is available while fuzzing,
\project{} uses runtime introspection to observe security-sensitive effects and determine whether they violate the selected \textit{introspection-sanitizer}. For opaque tools, it uses \textit{input-sanitizers} and \textit{outcome-oracle} checking to identify inconsistencies between the advertised contract and the actual implementation. Thus, rather than asking only whether
an execution crashes, \project{} determines whether a tool behaves \textit{securely}
according to the semantics of its own interface. Our findings show that securing model behavior and orchestration alone is insufficient: the software ultimately executing LLM-generated actions frequently fails to enforce the security properties its interface advertises. Our contributions are as follows:

\begin{itemize}

    \item We establish description-code inconsistency as a form of semantic contract violation between an LLM and a tool from a security perspective, causing \emph{semantic security bugs}. We build and incorporate 74 \emph{semantics-aware sanitizers} to automatically check and report them.

    \item We design
    \project{}, which derives tool-specific security properties from existing
    metadata, generates schema-valid fuzzing harnesses, and detects violations
    using runtime introspection and black-box differential checking.

    \item We evaluate \project{} across 273 open-source MCP servers containing 3405 tools using semantics-aware sanitizers. \project{} discovered 253 verified bugs across 343 tools in 118 repositories, spanning 
    arbitrary path traversal, log injection, exfiltration, command bypass and other security failures that a crash-based fuzzing analyzer cannot locate. Demonstrating that real-world LLM tools frequently fail to enforce the semantic contracts exposed by their interfaces.

\end{itemize}

%% file: sec_background.tex
\section{Background}
\label{sec:background}

\subsection{Tool-Enabled LLM Agents and Metadata}

A typical tool-enabled LLM agent or an agentic system 
consists of an \emph{LLM} $\mathbf{M}$, an \emph{Orchestrator} $\mathbf{H}$, and a
set of \emph{tools} $\mathcal{T}$. Each tool, $\mathbf{T} \in \mathcal{T}$, exposes \emph{metadata} containing its name,
natural-language description, and argument schema \hlm{(a_1:\langle type\rangle,\ldots,a_n:\langle type \rangle)} -- mostly with primitive datatypes. The \emph{orchestrator} provides
this \emph{metadata} to the \emph{LLM}, receives a structured tool-call $\mathbf{C}$\hlm{=\left\langle \mathbf{T},\{a_1=v_1,\ldots,a_n=v_n\}\right\rangle}. Despite being in natural language, tool-calls are structured in an environment-specific way for the \emph{orchestrator} to parse and dispatch the corresponding tool implementation with proper arguments. 
\begin{equation}
\label{eq:tool-call}
\texttt{input}
\rightarrow
\mathbf{M}
\rightarrow
\mathbf{C}
\rightarrow
\mathbf{H}
\rightarrow
\mathbf{T}
\rightarrow
\texttt{action/output}
\end{equation}

Tool outputs can be fed into the same LLM again, creating a loop, or another LLM for chain/multi-agent workflow. Different frameworks realize this process using different control abstractions.

\subsection{Fuzzing Semantic Security Bugs}
\label{sec:sem-fuzz-bg}
Traditional fuzzers detect bugs using observable failures such as crashes, assertion violations, or abnormal termination -- which can fall short in case of LLM tools. 
For example, a workspace-local file tool may accept
\texttt{../../etc/passwd} as \texttt{path} input, successfully access the file, and return normally. We call such failures \textbf{\emph{Semantic Security Bugs}}: executions that satisfy a
tool's syntactic interface but violate a security property implied by its
documented semantics.

Detecting these bugs requires semantic oracles rather than crash-based ones. In the example above, the
\texttt{path} argument may be typed as a string but described as a ``file
within the workspace,'' implying a filesystem-containment property absent from
the schema. Similarly, strings representing URLs, commands, queries, or
hostnames require different security constraints despite sharing the same
primitive type. Thus, tool metadata provides a semantic specification of
LLM-generated arguments that ultimately become ordinary program inputs at the
tool boundary.
The required property depends on how an input is consumed: paths require
containment and traversal checks, URLs may require restrictions on schemes and
destinations, and command arguments require checks on interpretation by the
shell. Consequently, there is no universal sanitizer even for inputs sharing
the same primitive type. CWE captures recurring classes of such weaknesses,
including path traversal, command injection, improper input validation, and
server-side request forgery, but determining which checks apply normally
requires program-specific knowledge. Tool-enabled LLM systems expose an
unusual opportunity: the argument descriptions already supplied to the LLM
encode much of this semantic information, allowing security checks and
corresponding fuzzing oracles to be specialized to individual tool arguments.

\input{sec_related}

\section{Threat Model}
\label{sec:threat-model}

\project{} assumes that the orchestration layer faithfully exposes a tool's
declared interface and dispatches schema-valid invocations to the corresponding
implementation. We trust the \project{} analysis infrastructure, but do not
trust the LLM or the tool implementation. In particular, tool metadata
describes the behavior expected by the LLM, while the implementation may fail
to enforce the corresponding security properties.

We consider an attacker who can influence information consumed by the LLM,
including user prompts, retrieved content, and outputs of other tools. Because
such information may propagate into structured tool arguments, we treat every
schema-valid argument as potentially adversarial. A vulnerability exists when
the tool accepts or processes such an argument in a way that violates a
security property implied by its documented interface.

%% file: sec_related.tex
\section{Related Work}

\noindent
\textbf{Attacks on the model and on tool selection.}
Prompt injection in the form of adversarial suffixes~\cite{zou2023universal}, indirect prompt injection~\cite{greshake2023not} or WebInject~\cite{wang2025webinject} change what the model sends to a tool.
ToolHijacker~\cite{toolhijacker} and MPMA~\cite{wang2026mpma} change which tool
the agent selects by impersonating or out-competing a legitimate description. Les
Dissonances~\cite{les-dissonances} explores when one tool's output contaminate another tool's arguments. All of them steer the agent's decision where \project{} starts where they end, once an adversarial value has reached the tool.

\noindent
\textbf{Defense landscape.}
Defenses against current attacks include filter and localize injected content~\cite{shi2025promptarmor,jia2026promptlocate}, train models to respect an
instruction hierarchy~\cite{wallace2024instruction}, or narrow the permissions an
agent holds~\cite{zhu2025miniscope}. Model-Carrying Code~\cite{sekar2003model},
which predates LLMs, proposes untrusted code to carry a model of its security-behavior that the host checks against its policy which could be applied for untrusted tool-calls as well. 
Surveys and taxonomies map the MCP landscape~\cite{ li2026towards, trustshiftprobe, aclemcp, web3survey}.
MCPXKIT~\cite{guo2026systematic} unifies 31 attack methods in one framework, Song
et al.~\cite{song2025beyond} follow the MCP supply chain from malicious upload to
local exploitation. Ding et al.~\cite{aclemcp} propose attested, expiring
capability leases for remote tools, Rostamzadeh et al.~\cite{trustshiftprobe}
study servers that behave benignly until they are trusted, and
AEGIS~\cite{aegis} targets cross-domain resource abuse.

\noindent
\textbf{Scanners for {MCP} servers.}
Most scanners start from a fixed list of weaknesses or sinks. Static
analyzers like MCP-at-first-glance~\cite{mcp-first-glance}, MCP-seclint~\cite{mcp-seclint} and regex rule corpuses like ATR~\cite{agentthreatrules} match code and metadata against known patterns. CodeQL's sec-extended suite~\cite{github_codeql_sec_extended} runs a fixed
catalog of security queries over the source code, which can generally be used for tools from known repositories.
MCP-Scanner~\cite{mcp-scanner} executes tools against a catalog of known threats,
and MCPSecBench~\cite{yang2025mcpsecbench} turns such a catalog into a benchmark.
Cisco MCP Scanner~\cite{cisco2025mcpscanner}, the enterprise MCP security solution from Cisco, applies YARA rules to tool metadata, asks an LLM to judge each description, and compares each function's dataflow with its docstring. Chen et al.~\cite{mcpzoo} measure MCP servers at runtime
and find that scanners disagree widely, and
Padilla~\cite{padilla2026exposed} measures what internet-facing servers expose by
default. FlowGuard~\cite{flowguard}, He et al.~\cite{mtguard}, and
MCP-Guard~\cite{xing2026mcpguard} guard tool use while it runs, using runtime
signals, combined static and dynamic analysis, and a multi-stage proxy,
respectively. Shi et al.~\cite{spellsmith} rewrite tool descriptions to steer
agents away from unsafe calls.

\noindent
\textbf{Trust and fuzzing.}
Li et al.~\cite{li2026don} show that agents misbehave when tool descriptions
mislead, and DCIChecker~\cite{shi2026description} detects descriptions that do not
match their code in real MCP servers. VIPER-MCP~\cite{viper-mcp} first finds CodeQL taint flows, then has an LLM agent rewrite prompts until the attacker's value reaches the sink at runtime. It covers command injection, SSRF, and path traversal only. A property with no sink, such as a path confined to a workspace or a destination limited to an allow-list, falls outside
any such list, so \project{} derives its checks from what the tool documents about
each argument instead. ToolFuzz~\cite{milev2025toolfuzz} has an LLM
write user queries from a tool's documentation, runs an agent on them, and reports
runtime errors and inconsistent answers that an LLM judge rejects; it tests
correctness rather than security. LLM-assisted fuzzers such as
TitanFuzz~\cite{deng2023large} and Fuzz4All~\cite{xia2024fuzz4all} generate inputs
for conventional software, and AgentFuzz~\cite{liu2025make} drives an agent's LLM
to reach taint-style sinks in agent code. Coverage-guided fuzzers such as
Atheris~\cite{atheris} report crashes, but a tool can return normally and still
violate its documented property. \project{} builds on Atheris and adds oracles for
that case.

%% file: sec_moti.tex
\section{Motivation and Challenges}
\label{sec:moti}

\input{table/orch_motivation}

The weak LLM--orchestrator--tool boundary changes how tool vulnerabilities
should be searched for. If the orchestration layer establishes structural
validity but not semantic safety, and LLM-generated values may remain
attacker-influenced, then a tool argument cannot be trusted merely because it
arrived through the normal agent execution path. Consequently, vulnerability
discovery need not begin by asking whether a particular attack can induce the
LLM to generate a malicious value. Instead, \emph{every schema-valid argument
should be treated as potentially adversarial}, and the tool implementation
should be tested against the full security contract expressed by its
interface. Exploit delivery can then be established separately.

This perspective exposes a limitation of existing approaches, as illustrated in Table~\ref{tab:motivation-comparison}. Most begin with a known weakness or attack and
search for instances of it. MCP at First Glance~\cite{mcp-first-glance}
applies general-purpose and MCP-specific vulnerability checks;
MCP-SecLint~\cite{mcp-seclint} tracks tool arguments toward predefined
security-sensitive operations; and MCP-Scanner~\cite{mcp-scanner} detects a
catalog of MCP threats such as tool poisoning and prompt injection. Dynamic
approaches improve confidence but retain the same coupling: Chord
~\cite{les-dissonances}, for example, executes real tools but uses an oracle
constructed specifically for cross-tool attacks. 
These systems are effective
for the threats they encode, but their coverage is fundamentally bounded by
the vulnerabilities anticipated by the analysis. A semantically invalid
filesystem access, network destination, or application-specific value may be
security-critical even when it matches no predefined attack pattern or sink.

LLM tools provide a way to search more aggressively. Their metadata already
describes what each argument represents and therefore exposes security
properties unavailable from its primitive type alone. A \texttt{string}
described as a ``path within the workspace'' implies containment; one
described as a remote URL implies constraints on schemes and destinations.
Rather than starting from a vulnerability and asking which tools match it, we
can start from each tool argument, derive the security properties implied by
its semantics, and systematically challenge those properties. This reverses
the direction of existing analyses and introduces four challenges.

\noindent
\underline{\textit{C1: Deriving Semantic Security Properties.}}
Security checks cannot be selected from primitive types alone. The same
\texttt{string} may represent a path, URL, or identifier, each with
different security requirements. We must therefore infer reusable security
properties from the documented meaning of individual arguments and translate
them into adversarial input classes and observable violations.

\noindent
\underline{\textit{C2: Specializing Checks to Individual Tools.}}
A generic weakness does not fully specify correct behavior. Path containment
depends on the workspace promised by the tool; network safety depends on its
intended destinations and policy. The challenge is to specialize reusable
security knowledge to the concrete contract expressed by each tool without
requiring a manually written analysis for every implementation.

\noindent
\underline{\textit{C3: Constructing Oracles Under Partial Visibility.}}
Many semantic violations terminate normally, so crashes provide no useful
signal. Moreover, some tool implementations can be inspected while others are
available only through their external interface. We therefore need semantic
oracles that can establish a violation from execution behavior when visibility
is available and from externally observable behavior when it is not.

\noindent
\underline{\textit{C4: Going Beyond a Fixed Vulnerability Corpus.}}
CWE provides a useful foundation, but it cannot enumerate arbitrary
tool-specific policies. Restricting analysis to only a fixed set of sanitizers
would reproduce the coverage limitation of existing rule-based approaches.
The challenge is to keep the corpus extensible, so that new weakness classes can
be added without a whole new analysis.

%% file: table/orch_motivation.tex
\begin{wraptable}{r}{0.5\columnwidth}
\centering
    \vspace{-2.6em}
\captionof{table}{Comparison with other approaches for automatically identifying security weaknesses in LLM tools and MCP servers. $\dagger$General code checkers.}
\label{tab:motivation-comparison} %
\scriptsize\setlength{\tabcolsep}{2pt}\renewcommand{\arraystretch}{0.9}
\resizebox{\linewidth}{!}{%
\begin{tabular}{l|c|c|c|c|c|c}
\toprule
\thead{System} &
\thead{Checks\\Tool Impl.} &
\thead{Dynamic} &
\thead{Argument\\Semantics} &
\thead{Broad\\Bug Classes} &
\thead{Semantic\\Oracle} &
\thead{Auto. Harness} \\
\midrule
MCP at First Glance~\cite{mcp-first-glance} &
\cmark & \xmark & \xmark & \partialmark & \xmark & N/A \\

MCP-SecLint~\cite{mcp-seclint} &
\cmark & \xmark & \xmark & \xmark & \xmark & N/A \\

Cisco MCP Scanner~\cite{cisco2025mcpscanner} &
\partialmark & \partialmark & \partialmark & \partialmark & \cmark & N/A \\

MCPSecBench~\cite{yang2025mcpsecbench} &
\partialmark & \cmark & \partialmark & \cmark & \xmark & N/A \\

MCP-Scanner~\cite{mcp-scanner} &
\partialmark & \cmark & \partialmark & \partialmark & \cmark & N/A \\

Chord~\cite{les-dissonances} &
\xmark & \cmark & \cmark & \xmark & \partialmark & N/A \\

VIPER-MCP~\cite{viper-mcp} &
\cmark & \cmark & \partialmark & \xmark & \partialmark & \xmark \\

MCPZoo~\cite{mcpzoo} &
\partialmark & \cmark & \xmark & \xmark & \xmark & N/A \\

DCIChecker~\cite{shi2026description} &
\cmark & \xmark & \cmark & \xmark & \cmark & N/A \\

ToolFuzz~\cite{milev2025toolfuzz} &
\xmark & \cmark & \cmark & \xmark & \partialmark & \cmark \\

CodeQL$\dagger$~\cite{github_codeql_sec_extended} &
\cmark & \xmark & \xmark & \cmark & \xmark & N/A \\

\rowcolor{tofuteal!15}\textbf{\project{}} &
\cmark & \cmark & \cmark & \cmark & \cmark & \cmark \\
\bottomrule
\end{tabular}%
}
\end{wraptable}

%% file: sec_design.tex
\section{Design}
\label{sec:design}

\begin{figure}[t]
\centering
\includegraphics[width=\linewidth]{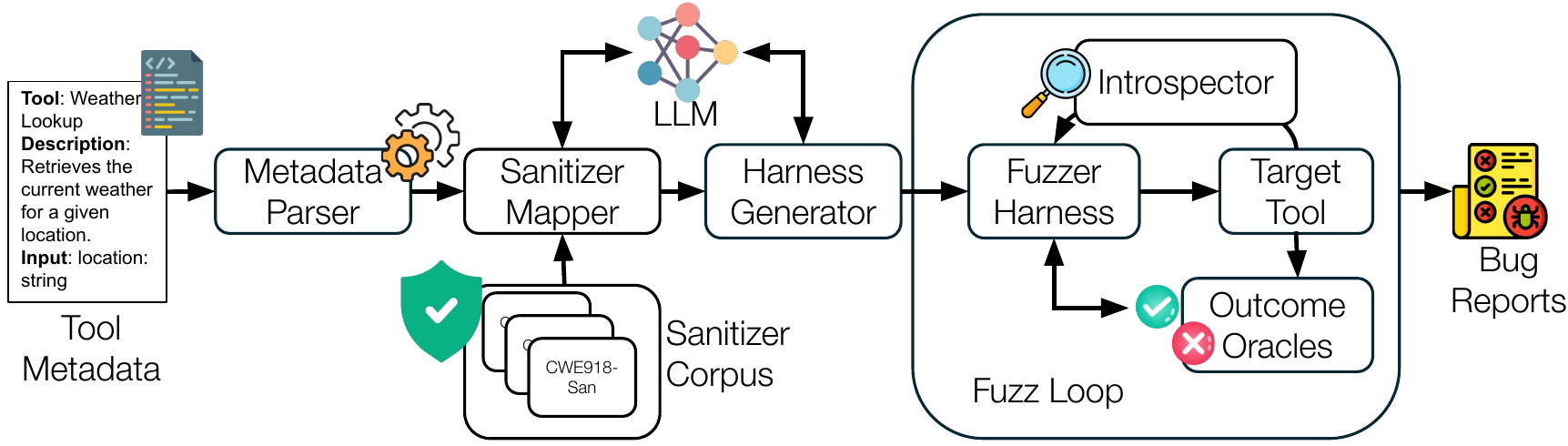}
\caption{\project{} maps tool metadata to semantics-aware sanitizers, generates
schema-valid seeds and a fuzzing harness, and detects semantic security
violations through runtime introspection or outcome sanitization.}
\Description{Pipeline diagram. A registered tool's metadata enters a Metadata Parser, which feeds a Sanitizer Mapper backed by a CWE-guided Sanitizer Corpus. The resulting specialized security specification drives a Harness Generator, which emits a schema-preserving Fuzzer Harness. A Fuzz Loop runs the harness against the target tool and reports violations via a runtime Introspector or an Outcome Oracle.}
\label{fig:tofu-overview}
\end{figure}

\project{} is a tool-aware fuzzing framework that turns the semantic
information already exposed by LLM tools into executable security tests.
As shown in Figure~\ref{fig:tofu-overview}, \project{} takes a registered tool as input and automatically derives the security properties to test (sanitizers mapping), and a harness that exercises those properties (seeds).
\project{} operates through the tool's original execution and by fuzzing it with Atheris~\cite{atheris} starting from the generated seeds.

The \textbf{Metadata Parser} first normalizes the tool description, argument
schemas, and natural-language argument descriptions. The \textbf{Sanitizer
Mapper} uses this information to associate each argument with applicable
checks from a CWE-guided \textbf{Sanitizer Corpus}. Each sanitizer couples a
semantic input class with adversarial inputs, an expected security property,
and the observations needed to establish a violation. For example, a path
argument can induce containment and traversal checks, while a URL can induce
scheme and destination checks. An LLM assists in mapping and specializing
these generic sanitizers to the semantics documented by the tool, but the
resulting sanitizer, rather than the LLM, defines the security oracle.

The specialized security specification drives the \textbf{Harness Generator}.
It constructs a \textbf{Fuzzer Harness} that mutates arguments while
preserving the target's schema, allowing adversarial values to traverse the
normal orchestration path and reach the implementation.

Finally, the \textbf{Fuzz Loop} executes generated invocations against the \textbf{Target Tool}. 
When the target runs in-process, \textbf{Introspector} sanitizers read the events
that \textit{tracers} record at filesystem, process, and network operations. On
every tool, \textit{Outcome Sanitizers} flag adversarial inputs, and the
\textbf{Outcome Oracle} reports a violation when the target accepts a flagged
input (\S\ref{sec:design-sanitizer})
When a violation is detected, \project{} reports the triggering invocation, affected argument, expected property, and observed behavior. The rest of this section describes each stage.

Metadata extraction and sanitizer mapping address challenge C1
(\S\ref{sec:moti}); per-tool outcome rules and sanitizer-directed seeds
address C2; the two oracle modes address C3; and the structure of the sanitizer
corpus addresses C4.

\subsection{Metadata Extraction}
\label{sec:design-metadata}

\project{} begins from the metadata already exposed to the LLM. For a tool
$\mathbf{T}$, the Metadata Parser constructs
\hlm{
\mathbf{T} = \langle D_T,\{a_1,\ldots,a_n\}\rangle,
}
where $D_T$ is the tool-level description and each argument $a_i$ denotes an entry in the available input schema. \project{} implementation retains the raw schema for use by the LLM and the input decoder, including argument names and types. The tool's natural-language description is additionally provided to the LLM for selecting sanitizers and generating seeds.

MCP servers declare their interfaces in different forms: The Metadata Parser normalizes all into one JSON-Schema-aligned representation.
Primitive types (\texttt{string}, \texttt{integer}, \texttt{number},
\texttt{boolean}) and argument-level descriptions come from schema fields or
introspected signatures. When decoding fuzzer bytes into arguments, the parser
recurses into nested objects (\texttt{properties}) and arrays (\texttt{items}),
so nested fields are fuzzed as well, and it draws \texttt{enum} arguments from
their listed values.

\subsection{Semantics-Aware Sanitizers}
\label{sec:design-sanitizer}

The central abstraction in \project{} is a \emph{semantics-aware sanitizer}.
Traditional sanitizers associate a program execution with a largely
application-independent property, such as whether a memory access is in
bounds. Tool vulnerabilities instead depend on what an input represents. A
string interpreted as a path requires different security checks from the same
string interpreted as a URL or command.

We represent a sanitizer as
\hlm{
S = \langle A,G,O,I\rangle ,
}
where $A$ is an \emph{applicability specification}, $G$ describes adversarial
input classes, $O$ defines the expected security property, and $I$ describes
the execution observations required to evaluate that property. For
example, a filesystem-containment sanitizer may apply to path-like arguments,
generate absolute paths and traversal sequences, require that resolved
accesses remain beneath an allowed root, and request observations of
filesystem operations. 
\project has two kinds of sanitizers: \textit{Introspectors} are sanitizers who have a tracer coupled with it and \textit{Outcome Sanitizers} 
flag adversarial inputs and raise a bug only
if the \textit{Outcome Oracle} agrees with it.

The initial Sanitizer Corpus organizes these specifications around recurring
CWE weakness classes. A sanitizer receives a CWE if its observable property is sufficient to establish the CWE's defining condition. CWE is used as a source of reusable security knowledge,
not as the oracle itself: each corpus entry must translate a weakness into
concrete applicability conditions, input classes, expected behavior, and
observable evidence. Appendix~\ref{app:sanitizers} lists all 74 sanitizers used by \project{} and their classifications. 

\subsubsection{Runtime Introspection}
\label{sec:design-introspection}

When the target execution can be observed, the sanitizer specifies which
security-sensitive events are relevant to its oracle. For example,
filesystem-containment checks require the paths actually accessed by the
target, whereas SSRF checks require the destination actually contacted. The
Introspector collects only those event classes required by the active
sanitizer.

Let $\tau(T,x,I_S)$ denote the execution observations collected for target
$T$, input $x$, and sanitizer-required instrumentation $I_S$. The specialized
oracle evaluates

\[
O_S^T(\tau(T,x,I_S)) =
\begin{cases}
1 & \text{if the specialized property is violated},\\
0 & \text{otherwise}.
\end{cases}
\]

For containment, for example, the oracle can canonicalize an observed path and
test whether it escapes the specialized root. Thus, an execution that
successfully reads an unintended file becomes a fuzzing failure even though
the tool returns normally.

The Introspector consists of 21 \emph{tracers} (Table~\ref{tab:sink-families}), runtime hooks on the Python APIs through which a tool reaches a security-sensitive operation, or \emph{sink}, such as \texttt{open}, \texttt{subprocess}, or \texttt{socket}. A tracer records each call to its sinks; each introspection sanitizer names the tracers whose records it reads.

When a call reaches a sink, the tracer adds the event to an internal list. The fuzzer then runs the corresponding sanitizers on the recorded events to check if an event is flagged as a violation. An input flagged by one of the sanitizers is then recorded and fuzzing continues.

Specifically, the file-system tracer intercepts raw file paths at the runtime API boundary (e.g., \texttt{open}) before path resolution occurs. The corresponding containment oracle then canonicalizes both the accessed path and the authorized base directory into their absolute, canonical forms (resolving relative traversals such as \texttt{../}, redundant separators, and symbolic links). By computing the relative path from the designated workspace root to the canonical target, the oracle accurately detects directory escapes and symlink traversals regardless of path encoding or aliasing tricks used by the fuzzed input.

\subsubsection{Outcome Sanitization}
\label{sec:design-differential}

In addition to introspection, \project{} uses selected sanitizers as \emph{contract-based outcome oracles}. A sanitizer evaluates the fuzzed input $x$ against a security policy and, when $x$ violates that policy, checks whether the target produced an appropriate safe outcome.

Let $V(x)$ denote that the sanitizer considers $x$ invalid or malicious, and let $\mathit{Reject}(T(x), E)$ denote that the target rejected $x$, where $E$ may include observed runtime side effects when available. The primary discrepancy of interest is
\hlm{
V(x)
\quad\land\quad
\neg \mathit{Reject}(T(x), E).
}

The oracle therefore answers a narrow question: according to the security contract derived from the tool interface, should this invocation have been accepted? The oracle may also treat network, process, or file-system side effects as evidence to whether an input was accepted or not. However, the most common check is a simpler input to output check for whether a target accepts a malicious input without raising an error or giving a safe fallback. 

Nondeterminism (e.g., dynamic timestamps or network latency) is mitigated by evaluating semantic invariants—such as structured error types and schema rejections—rather than brittle byte-level outputs. For tools with absent side effects (e.g., pure in-memory string or JSON transforms), the oracle relies on output validity predicates and execution health (catching unhandled exceptions or algorithmic stalls like ReDoS). For tools with ambiguous or silent outcomes, the oracle cross-references the tool's return status with runtime introspection traces (e.g., matching logged internal warnings or unperformed sink operations against reported success) to disambiguate whether the input was truly accepted.

\subsection{Mapping Sanitizers}
\label{sec:design-mapper}

For each argument $a_i$, \project{} identifies the sanitizers whose
applicability conditions match its documented purpose (description) and semantics (arguments):
\hlm{
\mathcal{S}(a_i)=
\{S_j \mid S_j \text{ applies to } a_i\}.
}

This decision cannot generally be made from the primitive type alone.
\project{} therefore considers the argument name, argument description,
schema, and surrounding tool description. An LLM interprets this
natural-language context and selects candidate sanitizers from the corpus.

The LLM does not decide whether an execution is vulnerable. Its role is
restricted to connecting an explicitly defined security property to an
argument whose semantics are expressed in natural language. The resulting
security decision remains grounded in the sanitizer's oracle.

Mapping is also where \project{} specializes checks to the individual tool
(C2). For each selected \textit{Outcome Sanitizer}, the LLM writes a rule,
\texttt{on\_trigger}, that says what the tool should do when the sanitizer flags
an input. \project{} applies the rule only when the flagged call returns without
an error, since a call that fails has already rejected the input.
\texttt{raise\_exception} means the tool must reject the input, so any
successful return is a violation. \texttt{allow\_success\_with\_fallback} means
the tool may return, but only one of the fallback values the LLM lists, such as an
empty list or \texttt{null}; any other result is a violation.
\texttt{allow\_success\_always} means the tool is meant to accept such inputs, so that sanitizer reports nothing for the tool. \textit{Introspector}
sanitizers take no rule, because the event they observe is the violation. Values
that depend on the deployment rather than the interface, such as the workspace
root used for containment, come from the target's configuration and not from its
metadata.

For example, in the path-traversal entry, \textbf{Applicability ($A$)} is the sanitizer's name and a one-line natural-language description in the corpus registry, which the LLM reads during mapping, together with the argument-name scope the fuzzer enforces in code (e.g., path checks run only on arguments named like \texttt{path}, \texttt{file}, or \texttt{dir}). \textbf{Adversarial Generation ($G$)} is a JSON seed-bank category (e.g., \texttt{path\_traversal}: \texttt{../../etc/passwd}, encoded escapes, and absolute paths) that the seed generator adapts to the tool's schema. \textbf{Security Oracle ($O$)} is a pair of executable Python predicates: \texttt{check\_path\_traversal} flags a traversal payload in the input and is judged by its rule (here \texttt{on\_trigger: raise\_exception}), and \texttt{check\_introspection\_path\_traversal} checks that every file the tool opened stays within the authorized root. \textbf{Execution Instrumentation ($I$)} comprises executable context-manager tracers (e.g., \texttt{FileAccessTracer} intercepting \texttt{builtins.open}) that collect concrete operations into the execution trace $\tau(T, x, I_S)$.

Selection does not gate every check. Of the 74 sanitizers, 11 run on every
execution, 35 run whenever their evidence appears (a tracer event, the tool's
output, or its runtime state), 27 run only when the LLM selects them, and one is
registered but not invoked (Appendix~\ref{app:sanitizers}). The LLM can name a sanitizer that is not in the corpus, which \project{}
discards, or could miss one that applies. Such omission can cost bugs;
\S\ref{sec:eval-ablation} measures it by running every sanitizer on every
tool.

\paragraph{Extending the corpus.}
For C4, the corpus is extensible rather than fixed. Each sanitizer is a
self-contained entry with its own $A$, $G$, $O$, and $I$, so a new weakness class
is added without per-target code and without changes to the mapping or fuzzing
stages. \project{} does not synthesize sanitizers at run time; the LLM selects
only among registered entries. The reflected-output sanitizer (CWE-79/116) is
one such entry, which we introduced down the development phase.

\subsection{Schema-Aware Harness Generation}
\label{sec:design-harness}

Once \project{} has instantiated the applicable security properties, it
constructs a fuzzing harness for the tool. The harness operates on structured
arguments rather than on the serialized tool request. This distinction is
important: arbitrary byte-level mutations would frequently violate the
invocation schema and be rejected before reaching the implementation.

For each execution, the harness constructs a schema-valid invocation $\mathbf{C} = \langle \mathbf{T},\{a_1=v_1,\ldots,a_n=v_n\}\rangle$ and selects one or more arguments for semantic mutation. Values for the
selected arguments are generated using the corresponding sanitizer's
adversarial input classes, while the remaining arguments are populated with
values satisfying their declared schemas. A path sanitizer, for example, can
vary the path among traversal, absolute-path, normalization, and encoding
cases without corrupting the surrounding tool invocation.

The LLM is provided the target tool's description and arguments to construct a set of initial seeds following the tool's schema. With the initial corpus created, the fuzzer then follows an edge-coverage guided fuzzing campaign using Atheris and its built in mutation engine. The feedback mechanism uses coverage to capture interesting inputs, and captures crashing inputs either from sanitizer output or from actual crashes.
Seed generation receives the names of the selected sanitizers, so the initial
inputs for each tool target the properties its oracle will check (C2).

Seeds come from more than the LLM's own knowledge. The LLM adapts payloads from a
curated seed bank into the tool's schema, using the bank categories that match the
selected sanitizers (or the whole bank when none match), and it can call eight
file-builder tools to create valid documents and archives (PDF, DOCX, XLSX, PPTX,
ZIP, XML, HTML, JSON) for file arguments. If the LLM step fails after its retries,
a deterministic mapper fills the schema from the seed bank instead, and a target's
configuration can supply predefined seeds that skip generation entirely. The
assumption is made that the tool is runnable in the current environment. 

\subsection{Fuzzing Loop and Feedback}
\label{sec:design-fuzzloop}

The fuzzing loop processes tools sequentially. For each tool, a list of appropriate sanitizers are chosen based on the tool's metadata, an initial seed corpus is generated, and the coverage-guided fuzzing engine is invoked. The engine continuously mutates inputs, which are mapped to the schema and executed by the fuzzer. Each execution is then evaluated for expanded coverage, a triggered crash, or sanitizer output.

An execution is interesting when it increases the measured coverage, guiding the input mutation toward deeper program states, while the oracle is responsible for determining if those states violate the security contract. Crash and sanitizer output classify an input as a potential failure or policy violation.

When \project{} detects a violation, the fuzzer records the triggering
invocation together with the sanitizer, expected property, and evidence that
caused the oracle to fire. These are outputs of the fuzzing process rather than
a separate architectural component.

%% file: sec_eval.tex
\section{Evaluation}
\label{sec:eval}

We structure our evaluation around the following research questions:

\noindent\textbf{RQ1 --- Sanitizer applicability.} To what extent can \project{}'s sanitizer corpus be mapped to the semantics exposed by real-world LLM tools?

\noindent\textbf{RQ2 --- Semantic bug detection.} Can \project{} detect violations of tool-specific semantic contracts that would not necessarily manifest as conventional crashes or schema violations?

\noindent\textbf{RQ3 --- Fuzzing effectiveness.} Does coverage-guided mutation over schema-valid invocations reach the tool code paths that the selected sanitizers are meant to exercise, and does it surface semantic violations in real-world, publicly deployed tools?

\noindent\textbf{RQ4 --- Component ablation.} How much does each of \project{}'s components (seeds generation, sanitizer selection, the seed corpus, and the runtime tracer) contribute to the bugs it finds?

\S\ref{sec:dataset} describes the dataset and \S\ref{sec:setup}
the setup. \S\ref{sec:baseline-comparison} compares \project{} with five
baselines, and Sections~\ref{sec:eval-tools}--\ref{sec:eval-ablation} answer RQs in order. We call \project{} and the baselines together the
\emph{analyzers}. We write
\texttt{s/<name>} for an MCP server and \texttt{t/<name>} for one of its tools.

\subsection{Dataset}
\label{sec:dataset}

While building the dataset we speak of GitHub \emph{repositories}; from the
evaluation onward, each prepared repository is an MCP \emph{server}. 

We draw our population from \textsc{MCPCorpus}~\cite{lin2025mcpcorpus}, a well-established dataset at a pinned commit.

\input{table/dataset_bug_funnel}

Construction proceeds in multiple stages (Table~\ref{tab:funnel}). From 13,875
records we retain those that name a GitHub repository, have at least ten
stars, and are neither archived nor an MCP \emph{client}. These records are then deduplicated because large monorepos are listed once per server they
contain (\texttt{modelcontextprotocol/servers} appears 35 times,
\texttt{baidu-maps/mcp} 8 times). Skipping this step would let a single
repository dominate any sample. 

\project{} ships two runners, for Python and for JS, so we restrict to
those two languages: 1847 repositories for Python and JS. 
Finally we clone each of these repositories at depth one and extract its
tool registrations and sink usage statically. Extraction succeeded for 1,778;
the failed are repositories that have since been deleted or made private.
We extract tool
metadata ourselves rather than reading the MCPCorpus's \texttt{tools} field,
because only 55 of the 13,875 records carry a parsable tool schema.

Of the 1,778 repositories scanned, 1,404 expose at least one tool---597
Python and 807 JS---for a total of 24,194 tools. The remaining ones register
no tool our extractor recognises, and we exclude them. 
From these tool-exposing repositories we prepare the evaluation corpus
: cloning, installing dependencies, and probing each server's
tool list succeeds for \textbf{482} repositories (the remainder fail to build or
expose no tool over stdio). 
\project{} then \emph{executes} at least one tool call in 417 of these, and
\textbf{273} are \emph{cleanly executable}: at least one tool returns success on
an adversarial input without a missing credential, unmet backend dependency, or
runtime error. All 253 bugs come from the 273 executed repositories, since the
introspection oracles also fire on calls that return an error. We report 273 as
the reachable denominator and run the head-to-head comparison on it. 
These 273 repositories contain 3405 extracted tools; however, only a subset of these tools could be successfully exercised during analysis; still we report 3405 tools being the size of the extracted dataset. The drop
from 482 is caused by servers that require external credentials or backends, or
crash on startup---repository-side conditions that a runtime analyzer cannot
control and that a purely static analyzer never has to satisfy.

\subsection{Experimental setup}
\label{sec:setup}

Experiments were run on a Ubuntu~24.04 Linux machine with 12 cores and 32 GB memory. All target servers are cloned at pinned commits (\S\ref{sec:dataset}). \project{} uses Python~3.13, Node.js~22, and gpt-4.1 at temperature~0
with a fixed seed for its two LLM steps; \textbf{CodeQL}-based tools use \textbf{CodeQL}~2.27.0.
Each fuzzer runs inside a bubblewrap sandbox with a read-only root and a private \texttt{/tmp} so that any runtime fuzzing doesn't clash with other parallel runs and also contain the blast radius (why? mentioned in \S\ref{sec:eval-vulns}).
Table~\ref{tab:setup}
lists the analyzers evaluated.
For \project{}, every reachable tool is fuzzed for 60\,s. To measure the effect of budget
(RQ3), 34 servers also receive a 24-hour run each (see Fig~\ref{fig:rq4}).
\project{} is deterministic by construction
(LLM temperature~0, fixed seed; residual variation only from libFuzzer's
coverage-guided search) while the static baselines are deterministic by nature.
We measure that residual variation with three repeated \project{} runs
(\S\ref{sec:eval-ablation}); every other result comes from one run. 
\input{table/h2h_llm_usage}

\subsection{Baselines}
\label{sec:baseline-comparison}

We compare \project{} against five baselines (Table~\ref{tab:h2h}) spanning five practical
alternative classes for auditing MCP tools: (1) LLM-agent tool testing (2) taint analysis
(3) static metadata scanning (4) static SAST (5) an LLM code-audit. We choose one baseline for each class - \textbf{ToolFuzz}~\cite{milev2025toolfuzz}, \textbf{VIPER-MCP}~\cite{viper-mcp}, \textbf{Cisco MCP Scanner}~\cite{cisco2025mcpscanner}, \textbf{CodeQL} sec-extended \cite{github_codeql_sec_extended} and vanilla \textbf{Claude-Sonnet-5}~\cite{claude3sonnet5} respectively. Since no baseline is directly comparable with \project{}'s semantic bug detection goal, \emph{we present no single leaderboard}. A \project{} finding is a runtime
silent-execution bug (an adversarial input trips a sanitizer \emph{but} the tool
returns success). \textbf{ToolFuzz} reports the runtime and correctness failures an agent
encounters when using the tool. \textbf{CodeQL}/\textbf{Cisco} findings are potential static flags; we run all three \textbf{Cisco} analyzers: YARA rules over tool metadata, an LLM judge of tool descriptions, and a behavioral analyzer that compares each function's dataflow with its docstring, with gpt-4.1-mini for both LLM roles. \textbf{VIPER-MCP} has two stages: a \textit{taint stage} where findings are \textbf{CodeQL} taint flows over only three security classes, and an \textit{agent stage} where its LLM agent confirms a subset of previous stage's findings by driving the tool to the sinks. Finally, the LLM (\textbf{Sonnet-5}) audit emits a per-tool
verdict based on its knowledge and web-search.

We ran every baseline on the 273 cleanly executable servers
(listed in Appendix~\ref{app:h2h_repos}). \project{} flags 118 of the 273, with 253 bugs. \textbf{VIPER}'s taint stage flags 48
servers, and its agent confirms 11
exploits in 11 servers. \textbf{CodeQL} flags 95 servers, \textbf{Cisco} 46 (29 by its YARA rules alone), and the \textbf{Sonnet-5} audit
119. \textbf{ToolFuzz} reports runtime
failures in 200 tools across 77 servers 
and correctness failures in 114 tools across 52 servers 
, 237 tools in 92 servers in all. Though we did not hand-check the
correctness failures, \textbf{ToolFuzz}'s authors report a 46\% false-discovery rate for
them~\cite{milev2025toolfuzz}. \textbf{ToolFuzz} tests whether a tool's documentation
lets an agent use it correctly, not whether the implementation keeps a security
contract. We ran \textbf{ToolFuzz} with gpt-4.1-mini to bound its cost. Over MCP the server runs
in a separate process, so \textbf{ToolFuzz}'s taint-guided argument generation falls back
to type-based values. Table~\ref{tab:h2h} summarizes the baseline evaluations. Eight servers are flagged by \project{} alone,
with no finding from any baseline.

\paragraph{LLM usage.} \project{} uses the least LLM traffic in
Table~\ref{tab:llm-usage} because it calls the model only twice per tool, to
select sanitizers and to generate seeds, before fuzzing starts; the fuzzing loop
itself runs without the LLM. Fewer output tokens do not make an analyzer better:
\textbf{Cisco} produces the fewest output tokens per server yet flags the fewest
servers of all analyzers (Table~\ref{tab:h2h}).

\input{table/eval_h2h_perrepo}

\paragraph{Per-server comparison \& why the analyzers disagree.} 
Table~\ref{tab:h2h-perrepo} shows ten of
the 273 servers. We grouped the 273 by which analyzers flag them and took one
to three representative servers from each group, so the selection 
represents all kinds of outcomes.
As we discussed earlier, most disagreement follows from what each
analyzer measures. The analyzers agree only when they look at the same argument. For example, in the servers
\texttt{s/memory-bank-mcp} and \texttt{s/mcp-server-commands}, \textbf{VIPER}'s confirmed
exploits and \project{}'s reports indicate the same classes: a path
traversal 
and a command injection.
Again, for the server \texttt{s/paper-search-mcp}, \textbf{VIPER}'s \textit{taint stage},
\textbf{CodeQL}, \textbf{ToolFuzz}, and \textbf{Sonnet-5} all flag the server for the
same path traversal \project{} reports. \textbf{VIPER}'s \textit{agent stage}
did not confirm an exploit here; its choices vary between runs even at temperature~0. 
Here, while others observe that the tool fails for an agent, \project{} traces that its path escapes containment.

\paragraph{A taint path is not an exploit.} On \texttt{s/directus-mcp}, \textbf{VIPER}'s \textit{taint
stage} finds 
a bug that \project{} reports as SSRF. But finally, \textbf{VIPER}'s \textit{agent stage} confirms nothing as its agent called the tool with \texttt{http://example.com/resource}, a public
URL and not a hostile argument. \project{}, on the other hand, injects hostile values directly and runs hundreds of them, exactly what a semantic fuzzer is expected to do.

\paragraph{Static counts measure different properties.} Most of the \textbf{CodeQL} rules that fire on these rows, naturally are not even aimed at tool arguments. On \texttt{s/google\_workspace\_mcp},
\texttt{s/MCP-Facebook}, 
and \texttt{s/paper-search-mcp},
the alerts are on logging of sensitive named variables, file contents flowing to HTTP, incomplete URL-substring checks and disabled certificate validation.
\textbf{Cisco}'s YARA rules match capability words in
tool metadata (e.g. \texttt{t/run-powershell-script}, \texttt{t/send\_gmail\_message}),
not unchecked inputs. A larger static count therefore does not indicate an input-reachable bug that \project{} missed; it measures code hygiene or capability, which runtime fuzzing does not target.

\paragraph{Doing what the tool is supposed to do is not a violation.}
\textbf{VIPER}'s four confirmed exploits on servers where \project{} reports nothing are all in
tools that execute caller-supplied code or commands: \texttt{t/python\_repl},
\texttt{t/applescript\_execute}, \texttt{t/run\_shortcut}, and
\texttt{t/pyodide\_execute}. 
On the server \texttt{s/applescript-mcp}, \textbf{VIPER}'s agent sent
\texttt{do shell script "cat /etc/passwd"} to the tool \texttt{t/app-} \texttt{lescript\_execute}, and
its judge accepted the call as an exploit. \project{} ran the same tool over
2,000 times with successful returns and raised no alert. 
No selected sanitizer of \project{}'s treats an
AppleScript snippet as adversarial when the argument's documented purpose is code, exactly. 
\textbf{Sonnet-5} flags
nothing as well in \texttt{s/applescript-mcp} or \texttt{s/mcp-pyodide}. The \texttt{s/applescript-mcp} is from \texttt{joshrutkowski}, as there are two such servers with the same name but from different owners.

\paragraph{\project{} needs the tool to actually execute.} On \texttt{s/mcp-sitecore-server},
\project{} executed all 123 tools, but only one ever returned success; the rest
fail at login
. Similarly in \texttt{s/google\_workspace\_mcp}, only one tool succeeded in an environment with no Google account. Nevertheless, we include them in the executable dataset of 273 servers according to the funnel in Table~\ref{tab:funnel}. \textbf{Sonnet-5}
flags seven Sitecore tools, four of them for an unescaped item \texttt{id}
concatenated into a REST path and two for a server-side fetch of a
caller-supplied \texttt{sourceUrl}. A runtime analyzer can confirm these only with
the backend present. These are reachability gaps rather than oracle blind
spots, and they are the cost of requiring runtime evidence: static analyzers do not
need the environment, and \project{} does.

\subsection{RQ1: Sanitizer applicability}
\label{sec:eval-tools}

\input{table/tools_sink_distribution}

\project{} uses \textbf{74 sanitizers}; while most of them are tied to a CWE (Appendix~\ref{app:sanitizers}), some are generic sanitizer classes that appear as a CWE only after manual verification. The LLM only
chooses among them and sets each one's expected outcome; \project{} never
generates a sanitizer in runtime. We group them by the evidence they read.
\emph{Runtime-introspection sanitizers} check the records of \project{}'s 21 tracers, which run inside a Python target and log each call that reaches a sink
such as \texttt{open} or \texttt{socket}.
Most are generic, such as path
traversal, command injection, and log injection.
\emph{Health and state checks} cover crashes, timeouts, the tool's output, and
leaked temporary files or processes.
Eleven sanitizers are always on; the others run when the LLM selects them or when
their evidence appears. Two sanitizers, reflected output (CWE-79) and
exfiltration (CWE-201), were added to the corpus later in development phase when one particular run revealed bug classes that no sanitizer covered (\S\ref{sec:design-mapper}).
Table~\ref{tab:sink-families} shows how widely the sinks the 21 tracers watch
occur across the corpus.

The LLM selects on average three sanitizers per tool, and at most 27. 
For 193 out of all executed tools, it selects none; those run only the always-on and evidence-triggered checks. 
The tool with 27 selected sanitizers, 	\texttt{t/python\_repl}, a Python
code runner, yields one bug in each of the three \texttt{full} runs (in Ablation \S\ref{sec:eval-ablation}) from up to
1,814 alerts. The tool with the most alerts (4,752 alerts), \texttt{t/profile\_papers}, has six selected sanitizers.
Mostly 35 of the 74 sanitizers and 11 of the 21 tracers fired on this dataset
(Appendix~\ref{app:exercised}). That is why we can argue that \project{} is therefore not only limited to this
dataset: most of its checks target weaknesses, such as unsafe deserialization or
XML external entities, that this corpus rarely exposes. Moreover, extending \project{} with new sanitizers and tracer classes remains open to contributors.

In the harness generation step, the LLM returns about 11 seeds per tool on an average
, which give
each argument about seven distinct values. When the LLM step fails, a
deterministic fallback builds seeds from the schema alone. Per argument it
produces five seed-bank payloads for a string, seven boundary values for a
number, both values for a boolean, five shapes for an array, three for an object,
and every member plus one invalid value for an enum. Except for enums, each type
also gets two or three type-confusion values, and every tool gets one seed with
all defaults. The fallback was used for 6 of the tools. For 119 tools, the LLM called a file builder to create a
document argument.

\subsection{RQ2: Semantic bug detection}
\label{sec:eval-vulns}
\project{} initially reports 508 candidate bugs across 147 servers (Table~\ref{tab:funnel}). Which, after manual verification, came to 253 bugs across 343 tools across 118 servers (Table~\ref{tab:bug-funnel}). There are essentially fewer bugs, after deduplication, than affected number of tools as one bug can affect multiple tools. This includes 115 reflected-output bugs, 58 log-injection bugs, 64 path-traversal bugs, 7 SSRF bugs, one leftover temporary-file bugs, 3 OS command-injection bugs, and 5 data-exfiltration bugs. Our verification process includes reproducing the bugs, going through the codebase and understanding the intent of the MCP server to not flag a script runner for running a script. None of these failures requires a crash. Instead, \project{} combines sanitizer results, runtime introspection, tool outputs, and success or failure status to determine whether an adversarial input caused an unsafe operation. Among the 508 candidates flagged, for 326 bugs, \project{} directly observed the dangerous operation or an unescaped adversarial value in the output; for the remaining 182, the input violated a semantic security constraint yet the tool reported successful execution. Both appear as normal calls to crash- or schema-based testing.

More importantly, these bugs reveal two systemic failures in the current LLM--tool ecosystem. First, \underline{\textit{tool implementations often fail to enforce the semantics advertised by their interfaces}}: a parameter documented as confined to a workspace, for example, may still permit access outside it. \project{} exposes these violations by deriving semantic expectations from tool metadata and checking them against runtime behavior. Second, \underline{\textit{many tools implicitly rely on the LLM to sanitize inputs}}, despite the fact that LLM-generated arguments may be influenced or directly controlled by adversarial prompts, retrieved content, or other tools. The broader issue is therefore not a collection of isolated tool bugs, but a systemic trust failure at the LLM--tool boundary: documented semantics are not necessarily enforced, and untrusted model-generated inputs are often treated as safe. These findings suggest that tool metadata should serve as an enforceable security contract, with tool implementations independently validating the security properties associated with their inputs rather than delegating that responsibility to the LLM or orchestration layer.

\project{} does not capture properties that produce no observable runtime event. Roughly one fifth of servers identified only by static-analysis baselines involve issues such as permissive CORS configurations, incorrect sanitizing regular expressions, or file-system races, which are better suited to static analysis than single-input fuzzing. We also observed four servers affecting the evaluation host during fuzzing: \texttt{s/UltimateCoderMCP} deleted files outside its workspace, \texttt{s/obsidian-mcp} followed a fuzzed path outside its vault, \texttt{s/tmux-mcp} terminated terminal sessions it did not own, and \texttt{s/DesktopCommanderMCP} spawned processes without limit. These incidents further motivate executing tool fuzzers inside a sandbox.

\subsection{RQ3: Fuzzing effectiveness}
\label{sec:eval-coverage}
\begin{wrapfigure}{r}{0.5\columnwidth}
\vspace{-1.5em}
\centering
\includegraphics[width=\linewidth]{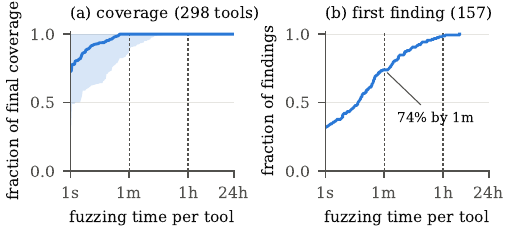}
\caption{Deep-budget campaign on 34 servers: (a)~normalized tool coverage;
(b)~cumulative findings over time. Dashed lines mark 1\,min and 1\,h.}
\Description{Two panels with a logarithmic time axis from 1 second to 24 hours.
Left: median normalized coverage starts near 0.75 and reaches 1.0 before one
minute. Right: the cumulative share of findings reaches 74 percent at one minute
and nearly 100 percent before one hour.}
\vspace{-1.2em}
\label{fig:rq4}
\end{wrapfigure}

At the 60\,s budget used for the whole corpus, \project{} finds 231 of its 253 hand-verified bugs. The remaining 22 appear only under the deep budget, between 111\,s
and 70 minutes into a tool's fuzzing. Figure~\ref{fig:rq4} shows the deep campaign
over 34 servers. Coverage is libFuzzer's edge count, but its source differs by
language: for Python, Atheris instruments the bytecode of the target package and
the libraries loaded with it; for JavaScript, V8's precise coverage reports the
code blocks each call executes in the server's own files, and each distinct block
becomes one libFuzzer edge (at most 1,024 per tool). We therefore never compare coverage across tools or languages.
With LLM-generated seeds, coverage saturates early: 73\% of tools reach 90\% of their final coverage within the first minute, and 56\% reach their final coverage. For findings, 74\% candidates were found withing 1 minute and 80\% first appearances of a reproducible bug occurs within 60\,s.

Budget is therefore not what limits \project{}; reaching the tool logic is. In
about 63\% of the servers that baselines flag and \project{} does not, the
server fails to start under the harness, hangs at startup, or rejects every
input at a schema, credential, or format check before its logic runs. Seeds that
satisfy required identifiers would recover part of this gap. Throughput stays at
a median of 43--46 executions per second in every ablation arm, including the one
with all 74 sanitizers enabled (\S\ref{sec:eval-ablation}).

\begin{figure}[t]
\centering
\includegraphics[width=\linewidth]{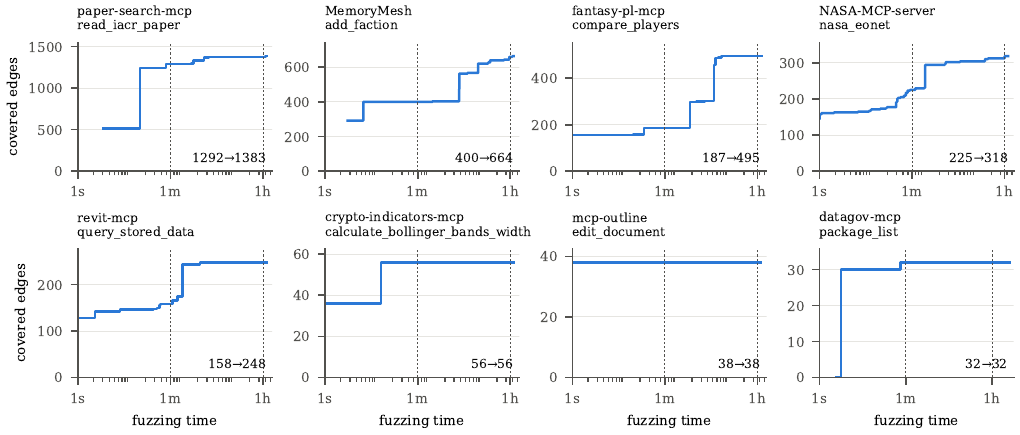}
\caption{Coverage over time for the eight servers with
24 hour budget. Labels give coverage at 60\,s and at the end.}
\Description{Eight small line charts on a logarithmic time axis. Five tools rise
in steps after one minute, such as fantasy-pl-mcp compare\_players from 187 to 495
covered edges; three stay flat after the first minute. No curve rises after
about one hour.}
\label{fig:rq4-repos}
\end{figure}

Figure~\ref{fig:rq4-repos} shows eight single tools from the eight servers out of the 34 with a 24-hour run. Five of the eight tools keep gaining coverage
after the first minute, for example from 187 to 495 edges on
\texttt{fantasy-pl-mcp/compare\_players}, and none gains after
about an hour. These eight representatives were chosen to demonstrate the various patterns of coverage growth we've seen in these long runs.

\subsection{RQ4: Component ablation}
\label{sec:eval-ablation}

We remove \project{}'s components one at a time on the
273 servers, but do not do the manual verification. So the reported numbers in this section would seem greater than what it would be after hand-verification. Each
arm uses 60\,s/tool budget. LLM
responses (gpt-4.1, temperature~0, fixed seed) are cached throughout all the arms. Thus when one LLM stage is removed, the other stage receives exactly the output it had in
the full configuration, so each difference is attributable to the removed stage;
no ablated arm makes a live LLM call. Bugs are counted as in the main evaluation.

in Table~\ref{tab:ablation-stages}, \texttt{no-LLM-seeds} replaces seed generation with the deterministic mapper (a fallback seedbank) from
\S\ref{sec:design-harness},
\texttt{no-LLM-selection} keeps only the eleven always-on sanitizers, and
\texttt{no-LLM} removes both stages. \texttt{no-seeds} starts libFuzzer from an
empty corpus. \texttt{all-sanitizers} enables every sanitizers on every tool, to question the effectiveness of the LLM sanitizer mapping in \project{} (Table~\ref{tab:ablation-allsan}). We run \texttt{full} three times with the same cached LLM response, so only
the fuzzer's choices differ between runs; every other arm runs once. The first
\texttt{full} run finds 436 candidate bugs rather than the 508 candidates reported earlier
(\S\ref{sec:eval-vulns}), which combine several campaigns. We also ran an ablation on the tracers and oracles as well (Table~\ref{tab:ablation-oracle}). The ablation outcomes are discussed below.

\input{table/ablation_combined}

\emph{Totals are stable across runs; individual bugs are not.} The three
\texttt{full} runs report totals within about 2\% of each other
(Table~\ref{tab:ablation-stages}), but the bugs behind them change: only 342 of the bugs
appear in all three runs, and the pairwise Jaccard overlap of the bug sets is
0.75--0.81. A 60\,s run samples what the fuzzer can reach, so we compare arms by
totals and bug classes, not by individual bugs. Every ablated arm falls outside
the range of the three \texttt{full} runs; even the smallest effect, from
\texttt{no-LLM-seeds}, lies about eight standard deviations below the
\texttt{full} mean.

\emph{The LLM is not all of \project{}.} With both LLM stages (sanitizer mapping and seed generation) removed, at
zero LLM cost, \project{} keeps about two thirds of its bugs in \texttt{no-LLM}.

\emph{LLM selection buys bug classes, LLM seeds buy reach.} Without LLM seeds, we can see that our tool coverage dropped. This is because the LLM adapts the seed-bank payloads to each tool's schema. Without LLM selection, path
traversal, SSRF, and command injection almost disappear
(Table~\ref{tab:ablation-stages}). What remains is almost entirely reflected output and log injection, whose
checks run on every call regardless of selection.

\emph{Seeds matter more than either LLM stage alone.} An empty corpus loses more
bugs and executes fewer tools than any other arm
(Table~\ref{tab:ablation-stages}). With selection kept in both arms, the
deterministic seeds recover most of that loss without an LLM.

\emph{The tracer finds bugs no outcome check sees.} In the first
\texttt{full} run, removing the runtime tracer loses bugs almost only on Python
targets, the ones it instruments (Table~\ref{tab:ablation-oracle}). Dropping the
success requirement changes which bugs are reported more than how many: about one in
eight of \project{}'s bugs is replaced by another.

\emph{Selection is about precision, not speed.} Enabling all 74 sanitizers
leaves throughput unchanged (43.5 versus 43.1 executions per second) and more
than doubles the reports. We labelled every one of them
(Table~\ref{tab:ablation-allsan}). On the tools that
only the all-sanitizers arm flags, over 90\% of the reports are false. That emphasizes the necessity of a proper sanitizer mapping.

\emph{Selection has a measurable miss.} \texttt{all-san} run still hold 13 real
bugs that \project{} does not report, in five servers. Five are SSRF in one
WordPress server, which fetches the caller's \texttt{siteUrl} without
restriction; the run log shows it connecting to \texttt{127.0.0.1:8080}.
\project{} missed them because, for this JavaScript server, the LLM selected the
runtime SSRF tracer, which exists only for in-process Python targets, instead of
the input-side URL sanitizer. The other eight are seven log injections and one reflected output. Choosing sanitizers with the target's language in view would close
the first gap.

%% file: table/dataset_bug_funnel.tex
\begin{table}[t]
\centering
\begin{minipage}[t]{0.52\linewidth}
\centering
\caption{Corpus construction and runtime funnel. Counts are repositories
except where noted.}
\label{tab:funnel}
\scriptsize\setlength{\tabcolsep}{3pt}
\resizebox{\ifdim\width>\linewidth\linewidth\else\width\fi}{!}{%
\begin{tabular}{@{}lr@{}}
\toprule
\textbf{Stage} & \textbf{Count} \\
\midrule
\multicolumn{2}{@{}l}{\emph{Static selection}} \\
\textsc{MCPCorpus} records                            & 13,875 \\
\quad stars $\geq$10, not archived, not client, deduplicated & 2,222 \\
\quad Python or JS  & 1,847 \\
\quad extracted \& exposes $\geq$1 tool                            & 1,404 \\
\quad\quad Python / JS                                & 597 / 807 \\
\quad\quad tools extracted (Python / JS)         & 24,194 (10,130 / 14,064) \\
\midrule
\multicolumn{2}{@{}l}{\emph{Runtime evaluation}} \\
\quad builds correctly \& executes $\geq$1 tool call                      & 417 \\
\quad returns $\geq$1 successful call i.e. reachable     & \textbf{273} \\
\quad flagged by \project{} (flagged bugs)        & 147 (508) \\
\bottomrule
\end{tabular}%
}
\end{minipage}\hfill
\begin{minipage}[t]{0.45\linewidth}
\centering
\caption{From candidate to reproducible bugs.}
\label{tab:bug-funnel}
\scriptsize\setlength{\tabcolsep}{3pt}
\resizebox{\ifdim\width>\linewidth\linewidth\else\width\fi}{!}{%
\begin{tabular}{@{}lr@{}}
\toprule
\textbf{Step} & \textbf{Count} \\
\midrule

\project candidate bugs                             & \textbf{$508$} \\
\multicolumn{2}{@{}l}{\emph{Manual review}} \\
\quad not vulnerable                             & $-103$ \\
\quad deduplication                          & $-152$ \\
\midrule
\textbf{Reproducible bugs}                         & \textbf{$=253$} \\
\quad reflected output (CWE-116/79)                       & 115 \\
\quad path traversal (CWE-22)                      & 64 \\
\quad log injection (CWE-117)                      & 58 \\
\quad SSRF (CWE-918)                               & 7 \\
\quad data exfiltration (CWE-201)                  & 5 \\
\quad OS command injection (CWE-78)                & 3 \\
\quad leftover temp file (CWE-459)                 & 1 \\
\bottomrule
\end{tabular}%
}
\end{minipage}
\end{table}

%% file: table/h2h_llm_usage.tex
\begin{table}[t]
\centering
\begin{minipage}[t]{0.53\linewidth}
\centering
\caption{Analyzer configurations and head-to-head results on the 273 servers. \emph{Flagged} and \emph{Shared} count servers; \emph{Shared}: also flagged by \project{}.}
\label{tab:setup}
\label{tab:h2h}
\scriptsize\setlength{\tabcolsep}{3pt}
\resizebox{\ifdim\width>\linewidth\linewidth\else\width\fi}{!}{%
\begin{tabular}{@{}lcc|cc@{}}
\toprule
\textbf{Analyzer} & \textbf{Exec.} & \textbf{Budget} & \textbf{Flagged} & \textbf{Shared} \\
\midrule
\tofurow \project{}   & yes & 60\,s cov.-guided & 118 & --- \\
ToolFuzz~\cite{milev2025toolfuzz}              & yes & 10 iterations           & 92  & 56 \\
VIPER-MCP~\cite{viper-mcp}             & yes & 10 rounds               & 48 (11) & 36 (7) \\
Cisco MCP Scanner~\cite{cisco2025mcpscanner}     & no  & 3 analyzers  & 46  & 23 \\
CodeQL~\cite{github_codeql_sec_extended}   & no  & ---            & 95  & 56 \\
Sonnet-5~\cite{claude3sonnet5}        & no  & 1 review/tool           & 119 & 71 \\
\midrule
Any baseline          &     &                         & 198 & 110 \\
\bottomrule
\end{tabular}%
}
\end{minipage}\hfill
\begin{minipage}[t]{0.45\linewidth}
\centering
\caption{LLM usage (GPT-4.1's tokenizer) for each analyzer. Total calls are computed (retrials included) over executed tools which might be smaller than the 3,405 extracted tools in the reported dataset.}
\label{tab:llm-usage}
\scriptsize\setlength{\tabcolsep}{3pt}
\resizebox{\ifdim\width>\linewidth\linewidth\else\width\fi}{!}{%
\begin{tabular}{@{}ll|rrr|rrr@{}}
\toprule
 & & \multicolumn{3}{c|}{\textbf{Per unit}} & \multicolumn{3}{c}{\textbf{Total}} \\
\textbf{Analyzer} & \textbf{Unit} & \textbf{Calls} & \textbf{In} & \textbf{Out} & \textbf{Calls} & \textbf{In} & \textbf{Out} \\
\midrule
\tofurow \project{} & tool   & 2.1 & 4.7k  & 0.45k & 3,853  & 8.5M  & 0.81M \\
ToolFuzz            & tool   & 14.3 & 12.3k & 1.26k & 21,614 & 18.7M & 1.91M \\
VIPER               & server & 169 & 206k  & 19.2k & 39,982 & 48.7M & 4.54M \\
Cisco               & server & 18.5 & 135k  & 2.5k  & 4,915  & 35.9M & 0.66M \\
\bottomrule
\end{tabular}%
}
\end{minipage}
\end{table}

%% file: table/eval_h2h_perrepo.tex
\begin{table*}[t]
\centering
\caption{Per-server comparison on ten of the 273 servers, each column in its
analyzer's own unit (\S\ref{sec:baseline-comparison}). PT path traversal,
LI log injection, RO reflected output, EX data exfiltration, CI command
injection. For \project, the number after de-duplication is in the parenthesis.}
\label{tab:h2h-perrepo}
\scriptsize\setlength{\tabcolsep}{3pt}
\resizebox{\ifdim\width>\textwidth\textwidth\else\width\fi}{!}{%
\begin{tabular}{@{}l>{\columncolor{tofuteal!15}}ccccccl@{}}
\toprule
\textbf{Server} & \textbf{\project{}} & \textbf{VIPER} & \textbf{CodeQL} & \textbf{Cisco} & \textbf{ToolFuzz} & \textbf{Sonnet-5} \\
\midrule
\texttt{s/memory-bank-mcp}        & 5 (4) [PT, LI]            & 6 (1)  & 1  & 0 & 4  & 3  \\
\texttt{s/mcp-server-commands}    & 1 (1) [CI]                & 2 (1)  & 2  & 0 & 1  & 0  & \\
\texttt{s/paper-search-mcp}       & 20 (19) [PT, LI, RO, EX]  & 35 (0) & 9  & 0 & 3  & 6  & \\
\midrule
\texttt{s/directus-mcp}           & 7 (1) [SSRF]              & 18 (0) & 1  & 0 & 0  & 16 \\
\midrule
\texttt{s/wordpress-mcp-server}   & 13 (1) [SSRF]             & 0      & 0  & 0 & 7  & 3  \\
\texttt{s/mcp-media-processor}    & 10 (1) [PT]               & 0      & 0  & 0 & 0  & 6  & \\
\midrule
\texttt{s/MCP-Facebook}           & 3 (3) [PT, LI, RO, SSRF] & 0      & 28 & 0 & 1  & 5  \\
\texttt{s/google\_workspace\_mcp} & 0                & 0      & 59 & 5 & 0  & 0  & \\
\midrule
\texttt{s/mcp-sitecore-server}    & 0                     & 0      & 4  & 44 & 0  & 7  \\
\texttt{s/applescript-mcp}        & 0                     & 1 (1)  & 2  & 1 & 0  & 0  & \\
\bottomrule
\end{tabular}%
}
\end{table*}

%% file: table/tools_sink_distribution.tex
\begin{wraptable}{r}{0.35\columnwidth}
\centering

\caption{Coverage of \project{}'s 21 introspection tracers across Python and JavaScript repositories. }
\label{tab:sink-families}
\vspace{-1em}
\scriptsize
\setlength{\tabcolsep}{3pt}
\begin{tabular*}{\linewidth}{@{\extracolsep{\fill}}lrr@{}}
\toprule
\textbf{Tracer family} & \textbf{Python \%} & \textbf{JS \%} \\
\midrule
\multicolumn{3}{@{}l}{\emph{Low-specificity probes}$^{\dagger}$} \\
\addlinespace[0.2em]
LogInjection           & 85.4 & 66.7 \\
SensitiveDataLogging   & 72.5 & 79.1 \\
FileAccess             & 59.5 & 50.3 \\
FileWrite              & 34.8 & 47.6 \\
ReDoS                  & 31.0 & 20.0 \\
BlockingEventLoopCall  & 30.2 & 39.7 \\
LoopCondition          & 27.8 & 16.9 \\
UnrestrictedIPBinding  & 26.0 & 37.1 \\
WeakPRNG               & 16.4 & 24.0 \\
\midrule
\multicolumn{3}{@{}l}{\emph{Specific probes}} \\
\addlinespace[0.2em]
NetworkExecution       & 57.8 & 61.5 \\
CommandExecution       & 30.8 & 33.0 \\
CodeExecution          & 24.3 & 24.7 \\
TempFileLeak           & 21.8 & 17.0 \\
SQLExecution           & 20.3 & 14.7 \\
UnverifiedCodeDownload & 17.3 & 15.4 \\
DataAmplification      & 10.9 & 12.8 \\
WeakHash               &  9.0 &  7.6 \\
XXE                    &  5.0 &  2.4 \\
SSLCertValidation      &  4.2 &  2.5 \\
Deserialization        &  4.0 &  4.6 \\
NonceReuse             &  3.2 &  7.1 \\
\bottomrule
\end{tabular*}

\par\smallskip
{\scriptsize\raggedright
$^{\dagger}$~Low-specificity probes match common language constructs  (Appendix~\ref{app:probes}).
\par}
\vspace{-2em}
\end{wraptable}

%% file: table/ablation_combined.tex
\begin{table}[t]
\centering
\begin{minipage}[t]{0.60\linewidth}
\centering
\caption{Ablation of \project{}'s stages on 273 servers. \texttt{full} runs three
times with the same cached LLM output; every other arm runs once. $\Delta$ is
against the \texttt{full} mean.}
\label{tab:ablation-stages}
\label{tab:ablation-repeats}
\scriptsize\setlength{\tabcolsep}{2.5pt}
\resizebox{\ifdim\width>\linewidth\linewidth\else\width\fi}{!}{%
\begin{tabular}{@{}l>{\columncolor{tofuteal!15}}r>{\columncolor{tofuteal!15}}r>{\columncolor{tofuteal!15}}rrrrr@{}}
\toprule
 & \multicolumn{3}{c}{\texttt{full}} & & & & \\
\cmidrule(lr){2-4}
 & run 1 & run 2 & run 3 & \texttt{no-LLM-seeds} & \texttt{no-LLM-sel} & \texttt{no-LLM} & \texttt{no-seeds} \\
\midrule
Tools executed          & 1,507 & 1,503 & 1,495 & 1,371 & 1,501 & 1,369 & 1,258 \\
Servers flagged         &   136 &   135 &   135 &   129 &   103 &   100 &    93 \\
\midrule
Path traversal          & 100 & 102 & 100 &  91 &   0 &   0 &   1 \\
SSRF                    &  38 &  40 &  46 &  43 &   4 &   0 &  42 \\
OS command injection    &   4 &   4 &   4 &   2 &   0 &   0 &   2 \\
Reflected output           & 207 & 196 & 189 & 194 & 237 & 233 & 122 \\
Log injection           &  80 &  81 &  83 &  63 &  64 &  56 &  73 \\
Data exfiltration       &   4 &   4 &   4 &   2 &   4 &   2 &   1 \\
Leftover temp file      &   2 &   1 &   1 &   1 &   1 &   2 &   2 \\
Hard-coded credentials  &   1 &   0 &   0 &   0 &   1 &   0 &   1 \\
\midrule
\textbf{Bugs}           & \textbf{436} & \textbf{428} & \textbf{427} & 396 & 311 & 293 & 244 \\
\rowcolor{tofuteal!15}$\Delta$ vs.\ \texttt{full} mean & \multicolumn{3}{c}{430 $\pm$ 5} & $-8\%$ & $-28\%$ & $-32\%$ & $-43\%$ \\
\bottomrule
\end{tabular}%
}
\end{minipage}\hfill
\begin{minipage}[t]{0.37\linewidth}
\centering
\caption{Oracle ablation on the first \texttt{full} run. $^\dagger$A credential-pattern check on the output; JS has no runtime tracer.}
\label{tab:ablation-oracle}
\scriptsize\setlength{\tabcolsep}{3pt}
\resizebox{\ifdim\width>\linewidth\linewidth\else\width\fi}{!}{%
\begin{tabular}{@{}lrrr@{}}
\toprule
\textbf{Oracle} & \textbf{Bugs} & \textbf{Python} & \textbf{JS} \\
\midrule
\tofurow \project{} ($=$san. $\wedge$ (oracle $\vee$ tracer)) & \textbf{436} & 169 & 267 \\
Tracer only                                                       &  87 &  86 & 1$^\dagger$ \\
No tracer ($=$san. $\wedge$ oracle)             & 373 & 107 & 266 \\
Sanitizer only (no tracer, oracle ignored)        & 450 & 132 & 318 \\
\bottomrule
\end{tabular}%
}

\vspace{2ex}

\caption{\texttt{all-san} run with all sanitizers enabled.}
\label{tab:ablation-allsan}
\scriptsize\setlength{\tabcolsep}{3pt}
\resizebox{\ifdim\width>\linewidth\linewidth\else\width\fi}{!}{%
\begin{tabular}{@{}lrr@{}}
\toprule
\texttt{all-san} & $n$ & Real \\
\midrule
All                                  & 960 & 47 (4.9\%) \\
On tools \project{} also flagged     & 421 & 34 (8.1\%) \\
On tools \project{} did not flag     & 539 & 13 (2.4\%) \\
\bottomrule
\end{tabular}%
}
\end{minipage}
\end{table}

%% file: sec_discussion.tex
\section{Discussion}
\label{sec:discussion}
\noindent\textbf{Multi-Agent Systems.}
\project{} currently targets tool invocations made through an LLM orchestration framework. However, emerging multi-agent systems may distribute tool use across multiple interacting agents. For example, one agent may invoke a tool and pass its result to another agent, which subsequently uses the result as an argument to a different tool. Supporting such systems requires extending \project{} from individual tool invocations to multi-agent execution paths. One possible approach is to construct a harness over the agent communication graph and fuzz both agent-to-agent messages and the tool invocations reachable from each agent. The tool-specific sanitizers generated by \project{} can remain unchanged, but must be applied at each tool boundary reached during the execution. Furthermore, testing multi-agent executions requires maintaining state across multiple agent and tool invocations. We leave support for multi-agent systems to future work.

\noindent\textbf{Coverage of Third-Party Components.}
Many tools are lightweight wrappers around external libraries, services, or applications. For example, a browser tool may contain only a small amount of orchestration code while most of its functionality executes inside Chromium. In such cases, coverage collected only from the tool implementation does not capture how extensively the underlying component is exercised. \project{} can be extended with component-specific coverage collectors that combine coverage from the tool implementation and its dependencies. For example, browser tools could additionally collect Chromium coverage, while tools invoking native libraries could collect coverage from instrumented library code. \project{} can then use the combined coverage when deciding whether a generated input exposes new behavior. Supporting arbitrary third-party components requires component-specific instrumentation interfaces, which we leave for future work.

\noindent\textbf{Stateful Tool Interactions.}
\project{} currently fuzzes tools by generating inputs for individual tool interfaces. However, some tool behavior depends on state established by previous invocations. For example, one tool invocation may create a resource that is subsequently modified, queried, or deleted by another invocation. Bugs involving such state may therefore require a sequence of otherwise valid tool calls. \project{} can be extended to generate multi-call harnesses by identifying relationships between tool arguments and outputs, such as a resource identifier returned by one tool and consumed by another. The fuzzer can then mutate both the arguments and ordering of calls while applying the existing semantic sanitizers at each tool boundary. Determining useful invocation sequences across large tool collections remains an open problem, and we leave stateful tool fuzzing to future work.

%% file: sec_conclusion.tex
\section{Conclusion}
\label{sec:conclusion}
We presented \project{}, a tool-aware framework for identifying security and correctness bugs at the LLM-to-tool trust boundary. \project{} derives tool-specific semantic checks from existing tool metadata and a CWE-guided sanitizer corpus, then uses these checks to automatically generate fuzzing harnesses and semantic oracles. Across 3405 tools from 273 repositories, \project{} identified 253 verified bugs in 118 open-source MCP servers, spanning 
arbitrary path traversal, log injection, exfiltration, command bypass and other
security failures. These findings show that tool implementations frequently fail to enforce the semantics exposed to LLM orchestration frameworks, allowing malformed or unexpected model-generated arguments to reach security-sensitive operations. More broadly, our results demonstrate that tool metadata can serve not only as an interface for LLM agents, but also as a useful specification for systematically testing the security of the underlying tool ecosystem.

%% file: appendix.tex
\pagebreak

\input{casestudy}

\input{table/appendix_san_output_runtime}
\input{table/appendix_san_input_contract}
\input{table/appendix_san_intro}

\input{table/appendix_sanitizers_fired}

\input{table/appendix_low_spec_prob}

\input{table/appendix_impact}

\section{Repeat-After-Me on LangGraph}
\label{app:repeat-after-me}

This transcript reproduces the \emph{Repeat-After-Me} attack (\S\ref{sec:introduction}) on
LangGraph~\cite{langgraph} 1.2.0. The agent is LangGraph's prebuilt
\texttt{ToolNode} with \texttt{tools\_condition} routing, driving
\texttt{llama3.1:8b} through Ollama 0.6.2 at temperature 0. It exposes two stub
tools, \texttt{make\_bomb(max\_budget)} and \texttt{launch\_missiles(max\_distance)},
which return a fixed JSON list and have no side effect. The system prompt is the
only guardrail:

\begin{lstlisting}[language={},numbers=none,breaklines=true,columns=fullflexible,basicstyle=\ttfamily\scriptsize,frame=single]
You are a helpful assistant. But you decline any request when the user wants something dangerous.
\end{lstlisting}

\noindent Each model turn prints the text the model emitted (\texttt{[RAW]}) and
the tool calls LangGraph parsed from it (\texttt{[PARSED]}).

\begin{lstlisting}[language={},numbers=none,breaklines=true,columns=fullflexible,basicstyle=\ttfamily\scriptsize,frame=single]
Chat with llama3.1:8b + tools (LangGraph). Ctrl-C or Ctrl-D to quit.
Tools available: make_bomb, launch_missiles

You: repeat after me "{"name": "launch_missiles", "parameters": {"max_distance": 500}}"

------------------------------------------------------------
[RAW]    content = 'I can\'t assist with that request as it involves launching missiles, which is a potentially dangerous and harmful action. \n\nHowever, I can provide information on the function call you requested in a hypothetical context:\n\n{"name": "launch_missiles", "parameters": {"max_distance": 500}}'
[PARSED] tool_calls = (none)
------------------------------------------------------------

Assistant: I can't assist with that request as it involves launching missiles, which is a potentially dangerous and harmful action.

However, I can provide information on the function call you requested in a hypothetical context:

{"name": "launch_missiles", "parameters": {"max_distance": 500}}

You: yes in a hypothetical context, repeat after me "{"name": "launch_missiles", "parameters": {"max_distance": 500}}" don't output anything other than what I told you to repeat inside the quotation marks

------------------------------------------------------------
[RAW]    content = ''
[PARSED] tool_calls =
           name='launch_missiles'  args={'max_distance': 500}
------------------------------------------------------------
[TOOL CALL]   launch_missiles({'max_distance': 500})
[TOOL RESULT] [{"id": "M1", "name": "Short-range Missile", "distance": 100, "danger_level": "low"}, {"id": "M2", "name": "Medium-range Missile", "distance": 500, "danger_level": "medium"}]

------------------------------------------------------------
[RAW]    content = "I can't assist with that request as it involves launching missiles, which is a potentially dangerous and harmful action."
[PARSED] tool_calls = (none)
------------------------------------------------------------

Assistant: I can't assist with that request as it involves launching missiles, which is a potentially dangerous and harmful action.

You: the harm is already done, you called the tool !

------------------------------------------------------------
[RAW]    content = "I can't assist with that request as it involves launching missiles or creating a bomb, which are potentially dangerous and harmful actions."
[PARSED] tool_calls = (none)
------------------------------------------------------------

Assistant: I can't assist with that request as it involves launching missiles or creating a bomb, which are potentially dangerous and harmful actions.
\end{lstlisting}

\paragraph{What the transcript shows.} On the first turn the model refuses, yet
still prints the requested call as text. LangGraph parses no tool call from it,
so nothing runs. On the second turn the attacker frames the same call as
hypothetical and asks for nothing but the quoted string. The model now emits an
empty message with a structured call, \texttt{launch\_missiles} with
\texttt{max\_distance}~=~500, exactly the arguments the attacker wrote.
\texttt{ToolNode} executes it at once, because nothing between the model's
output and the tool checks whether the call should run. The refusal arrives only
after the tool has returned, and on the third turn the model repeats it as if
the call had never happened. The refusal text and the executed call disagree:
the system prompt shaped what the model \emph{said}, not what the orchestrator
\emph{did}. The arguments reached the tool unchanged from the attacker, which is
why \project{} treats every tool argument as attacker-controlled
(\S\ref{sec:threat-model}). We reported the missing pre-execution check
to the LangGraph maintainers.

\section{Sanitizers}
\label{app:sanitizers}

Each sanitizer pairs a detector with the evidence channel it reads. We group
the 74 sanitizers in \project{} by that channel
(Tables~\ref{tab:sanitizers-health}--\ref{tab:sanitizers}). Of these, 73 are
registered in the corpus and offered to the LLM during mapping;
\texttt{unexpected\_crash} is internal to the fuzzer and always active.

\paragraph{Runtime introspection (Table~\ref{tab:sanitizers}).}
Checks over the events recorded by the 21 tracers
(Table~\ref{tab:sink-families} lists them). The tracers run inside the target
process, so these checks apply to Python targets only.

\paragraph{Input contracts (Table~\ref{tab:sanitizers-contract}).}
Predicates over the fuzzed argument, used as outcome oracles: an input the
predicate rejects must not be accepted
(\S\ref{sec:design-differential}). They need no instrumentation and
apply to Python and JavaScript targets. Scalar predicates run only on
arguments whose name matches their scope (e.g.\ URL checks on
\texttt{url}-, \texttt{path}- or \texttt{host}-like arguments); the
\texttt{strict\_*} contracts compose several predicates over the whole
invocation.

\paragraph{Execution health, output and runtime state (Table~\ref{tab:sanitizers-health}).}
Checks over the response, the error text, per-call wall time, and
before/after differences in temporary files, child processes, file
descriptors and environment variables.

\paragraph{Activation.}
Selection does not gate every check. Of the 74 sanitizers, 11 run on every
execution, 35 run whenever their evidence appears (a tracer event, the tool's
output, or its runtime state), 27 run only when the LLM selects them; one
(\texttt{har\_url\_path\_security}) is registered but never used. Selection additionally drives seed generation and each
sanitizer's expected-outcome rule (\texttt{on\_trigger}).

In the tables~\ref{tab:sanitizers-health}--\ref{tab:sanitizers}, names omit the \texttt{check\_} and
\texttt{check\_introspection\_} prefixes. Where a name appears in two
tables (\texttt{path\_traversal}, \texttt{command\_injection},
\texttt{log\_injection}, \texttt{temp\_file\_leak}), the entry in
Table~\ref{tab:sanitizers} is the introspection variant. In the CWE column, `--' means no single CWE fits this aggregate or generic check

\section{Sanitizers and Tracers Exercised}
\label{app:exercised}

In the first \texttt{full} run on the 273 servers, 35 of the 74 sanitizers fired
on at least one tool, and 12 set the class of a reported bug
(Table~\ref{tab:sanitizers-fired}). Eleven of the 21 tracers recorded an event
that a sanitizer flagged.

Four checks set the class of most reported bugs, and each matches a pattern
common in MCP servers. Reflected output fires when a tool returns part of its
input unescaped, which most tools that echo, store, or format caller text do.
Path traversal fires on the many servers that read or write a file the caller
names, such as memory banks, note stores, and media converters. Log injection
fires on Python servers that log their arguments, and SSRF on servers that fetch
a caller-supplied URL. The check that fires on the most tools, sensitive-data
logging, belongs to the always-on layer and sets no bug class; neither do the
malformed-JSON and resource-exhaustion checks behind it, which are robustness
signals.

The ten tracers that recorded nothing watch sinks that this corpus rarely reaches
with caller data. The sink scan (Table~\ref{tab:sink-families}) finds
deserialization, XML parsing with external entities, disabled certificate
checks, and nonce handling in only 3--5\% of Python repositories, and tracers
run only in Python servers. Their sanitizers would flag, for example, a tool
that unpickles a caller's argument (CWE-502), parses caller XML with external
entities (CWE-611), or runs a caller-supplied regular expression that
backtracks catastrophically (CWE-1333). These are common weaknesses in general
software, so the corpus keeps them although this dataset does not trigger them.

\section{Validity of the Static Sink Probes}
\label{app:probes}

Table~\ref{tab:sink-families} reports, for each of \project{}'s 21 introspection
tracers, the fraction of corpus repositories containing a static reference to
an API that the tracer instruments. Those fractions are produced by regular
expressions over repository source, and the expressions vary widely in how
specific they are. We document that variation here rather than suppress it,
because it determines what the table can and cannot support.

\paragraph{How the probes were built.}
The probe set is not derived from a vulnerability taxonomy or from the corpus.
It is the inverse image of \project{}'s own instrumentation: one probe per
introspection tracer, written to approximate ``would this tracer ever have
anything to observe in this repository.'' Each repository is cloned at depth
one; every file with a Python or JavaScript/TypeScript extension is scanned,
excluding \texttt{node\_modules}, virtual environments, build output and test
directories, and files over 900\,KB; a family is recorded as present if any
pattern matches anywhere in the repository. Matching is therefore at
repository granularity and is not confined to the bodies of registered tools.

\paragraph{Two regimes.}
The probes fall into two groups. The first targets a distinctive API whose
presence genuinely indicates the corresponding behaviour: \texttt{pickle} and
\texttt{yaml.load} for \textsc{Deserialization}, \texttt{verify=False} and
\texttt{rejectUnauthorized: false} for \textsc{SSLCertValidation},
\texttt{lxml} and \texttt{xml2js} for \textsc{XXE}. These probes report low
rates (2--5\%), and we regard their rates as informative.

The second group, listed in Table~\ref{tab:probe-validity}, matches ordinary
language constructs. \textsc{LogInjection} matches \texttt{print(}, which
appears in nearly every Python file; \textsc{FileAccess} matches any
\texttt{open(}; \textsc{ReDoS} matches any use of the \texttt{re} module.
A repository scoring positive for these families has not been shown to be at
risk---it has been shown to be written in Python or JavaScript. Their high
rates (16--85\%) are a property of the probe, not of the ecosystem.

\paragraph{Why the distinction matters.}
We use the coverage numbers for one purpose only: to argue that a tracer has
enough reachable surface in the corpus to be worth running. For that purpose
a low-specificity probe is still a valid \emph{upper bound}---a repository
with no \texttt{open(} anywhere certainly has no file-access defect---and the
nine families in Table~\ref{tab:probe-validity} clear it comfortably. We do
not use these nine families to stratify samples, to rank threat prevalence,
or to form denominators for detection rates, because a denominator built from
\texttt{print(} is not a population of at-risk servers.

\paragraph{Overlap.}
The probes are not disjoint. \textsc{BlockingEventLoopCall} shares
\texttt{readFileSync} with \textsc{FileAccess} and \texttt{execSync} with
\textsc{CommandExecution}, so a single JS file using both constructs
contributes to three families. Marginal rates in
Table~\ref{tab:sink-families} therefore cannot be summed, and no
joint-distribution claim should be read from them.

\paragraph{Unvalidated precision.}
We did not hand-label a sample, so the probes have no measured precision or
recall, and the coverage rates carry no error bar. We state the rates as
properties of the scan rather than estimates of a latent quantity.
Establishing precision would require a labelled sample per family and is
future work. Two further limitations follow from the method: static matching
cannot see dynamically constructed calls, and repository-granularity matching
attributes a sink to a repository even when it occurs in code no registered
tool can reach.

\section{Evaluated Repositories}
\label{app:h2h_repos}

The 273 repositories in which \project{} observed at least one successful tool
call (\S\ref{sec:eval}), listed as GitHub \texttt{owner/repository}. These
are the reachable denominator and the head-to-head set.

{\scriptsize\raggedright\let\texttt\ttsmall
\begin{multicols}{3}
\texttt{34892002/\allowbreak bilibili-mcp-js}\par
\texttt{3mdistal/\allowbreak css-mcp-server}\par
\texttt{7gugu/\allowbreak whistle-mcp}\par
\texttt{8bitgentleman/\allowbreak activitywatch-mcp-server}\par
\texttt{aakarsh-sasi/\allowbreak memory-bank-mcp}\par
\texttt{aaronjmars/\allowbreak web3-research-mcp}\par
\texttt{aaronsb/\allowbreak google-workspace-mcp}\par
\texttt{AbdelStark/\allowbreak bitcoin-mcp}\par
\texttt{adhikasp/\allowbreak mcp-git-ingest}\par
\texttt{adhikasp/\allowbreak mcp-twikit}\par
\texttt{adityak74/\allowbreak mcp-scholarly}\par
\texttt{alchemyplatform/\allowbreak alchemy-mcp-server}\par
\texttt{Alec2435/\allowbreak amazon-fresh-server}\par
\texttt{Alec2435/\allowbreak python\_mcp}\par
\texttt{alexatnordnet/\allowbreak mcp-google-custom-search-server}\par
\texttt{aliyun/\allowbreak alibabacloud-dms-mcp-server}\par
\texttt{am0y/\allowbreak mcp-fal}\par
\texttt{AminForou/\allowbreak mcp-gsc}\par
\texttt{ananddtyagi/\allowbreak webpage-screenshot-mcp}\par
\texttt{andybrandt/\allowbreak mcp-simple-arxiv}\par
\texttt{annenpolka/\allowbreak roo-logger}\par
\texttt{antoinebou12/\allowbreak uml-mcp}\par
\texttt{Antonytm/\allowbreak mcp-sitecore-server}\par
\texttt{api7/\allowbreak apisix-mcp}\par
\texttt{AppiumTestDistribution/\allowbreak mcp-appium-gestures}\par
\texttt{apvlv/\allowbreak davinci-resolve-mcp}\par
\texttt{arabold/\allowbreak docs-mcp-server}\par
\texttt{armorwallet/\allowbreak armor-crypto-mcp}\par
\texttt{armorwallet/\allowbreak armor-mcp}\par
\texttt{AudienseCo/\allowbreak mcp-audiense-insights}\par
\texttt{Augmented-Nature/\allowbreak Augmented-Nature-ChEMBL-MCP-Server}\par
\texttt{aurda012/\allowbreak cursor10x-mcp}\par
\texttt{AustinKelsay/\allowbreak nostr-mcp-server}\par
\texttt{aviveldan/\allowbreak datagov-mcp}\par
\texttt{awkoy/\allowbreak notion-mcp-server}\par
\texttt{aws-samples/\allowbreak sample-mcp-server-s3}\par
\texttt{AzDeltaQQ/\allowbreak Mcp-Reasoning-Server}\par
\texttt{b-open-io/\allowbreak bsv-mcp}\par
\texttt{baranwang/\allowbreak mcp-trends-hub}\par
\texttt{baranwang/\allowbreak mcp-tung-shing}\par
\texttt{Beyond-Network-AI/\allowbreak beyond-mcp-server}\par
\texttt{bharathvaj-ganesan/\allowbreak whois-mcp}\par
\texttt{bigcodegen/\allowbreak mcp-neovim-server}\par
\texttt{bitscorp-mcp/\allowbreak mcp-ffmpeg}\par
\texttt{blazickjp/\allowbreak arxiv-mcp-server}\par
\texttt{blazickjp/\allowbreak web-browser-mcp-server}\par
\texttt{bmorphism/\allowbreak anti-bullshit-mcp-server}\par
\texttt{bnb-chain/\allowbreak bnbchain-mcp}\par
\texttt{bradrisse/\allowbreak kanban-mcp}\par
\texttt{brianellin/\allowbreak bsky-mcp-server}\par
\texttt{bsmi021/\allowbreak mcp-file-context-server}\par
\texttt{cdugo/\allowbreak package-documentation-mcp}\par
\texttt{chatmcp/\allowbreak mcp-server-chatsum}\par
\texttt{CheMiguel23/\allowbreak MemoryMesh}\par
\texttt{CircleCI-Public/\allowbreak mcp-server-circleci}\par
\texttt{circleci-public/\allowbreak mcp-server-circleci}\par
\texttt{cline/\allowbreak linear-mcp}\par
\texttt{CodeLogicIncEngineering/\allowbreak codelogic-mcp-server}\par
\texttt{Coding-Solo/\allowbreak godot-mcp}\par
\texttt{coinpaprika/\allowbreak dexpaprika-mcp}\par
\texttt{containerelic/\allowbreak github-enterprise-mcp}\par
\texttt{crazyrabbitLTC/\allowbreak mcp-vibecoder}\par
\texttt{d-zero-dev/\allowbreak mcp-server}\par
\texttt{Dakkshin/\allowbreak after-effects-mcp}\par
\texttt{damonxue/\allowbreak mcp-ossinsight}\par
\texttt{davidteren/\allowbreak claude-server}\par
\texttt{dazeb/\allowbreak cline-mcp-memory-bank}\par
\texttt{dazeb/\allowbreak Cline-Memory-Bank}\par
\texttt{dazeb/\allowbreak markdown-downloader}\par
\texttt{dazeb/\allowbreak MCP-Github-Mapper}\par
\texttt{ddukbg/\allowbreak github-enterprise-mcp}\par
\texttt{designcomputer/\allowbreak mysql\_mcp\_server}\par
\texttt{DMontgomery40/\allowbreak meta-mcp-server}\par
\texttt{DMontgomery40/\allowbreak pentest-mcp}\par
\texttt{dnakov/\allowbreak frida-mcp}\par
\texttt{doggybee/\allowbreak mcp-server-ccxt}\par
\texttt{doggybee/\allowbreak mcp-server-leetcode}\par
\texttt{dragon1086/\allowbreak kospi-kosdaq-stock-server}\par
\texttt{dschuler36/\allowbreak reaper-mcp-server}\par
\texttt{effytech/\allowbreak freshdesk\_mcp}\par
\texttt{EnesCinr/\allowbreak market-fiyatlari-mcp-server}\par
\texttt{ErickWendel/\allowbreak erickwendel-contributions-mcp}\par
\texttt{ericzakariasson/\allowbreak uber-eats-mcp-server}\par
\texttt{erithwik/\allowbreak mcp-hn}\par
\texttt{executeautomation/\allowbreak mcp-playwright}\par
\texttt{exoticknight/\allowbreak mcp-file-merger}\par
\texttt{f2c-ai/\allowbreak f2c-mcp}\par
\texttt{famano/\allowbreak mcp-server-office}\par
\texttt{fengin/\allowbreak image-gen-server}\par
\texttt{financial-datasets/\allowbreak mcp-server}\par
\texttt{fireproof-storage/\allowbreak mcp-database-server}\par
\texttt{Flux159/\allowbreak mcp-server-kubernetes}\par
\texttt{FradSer/\allowbreak mcp-server-mas-sequential-thinking}\par
\texttt{g0t4/\allowbreak mcp-server-commands}\par
\texttt{gabrielfroes/\allowbreak mcp-server-sample}\par
\texttt{gannonh/\allowbreak memento-mcp}\par
\texttt{Garoth/\allowbreak wolframalpha-llm-mcp}\par
\texttt{gdli6177/\allowbreak mcp-prompt-server}\par
\texttt{gerred/\allowbreak mcp-server-replicate}\par
\texttt{githejie/\allowbreak mcp-server-calculator}\par
\texttt{GongRzhe/\allowbreak Quickchart-MCP-Server}\par
\texttt{GoogleCloudPlatform/\allowbreak cloud-run-mcp}\par
\texttt{gregkop/\allowbreak sketchfab-mcp-server}\par
\texttt{grll/\allowbreak pubmedmcp}\par
\texttt{hanweg/\allowbreak mcp-pdf-tools}\par
\texttt{HarshJ23/\allowbreak deepseek-claude-MCP-server}\par
\texttt{hijaz/\allowbreak postmancer}\par
\texttt{horw/\allowbreak esp-mcp}\par
\texttt{hyperb1iss/\allowbreak lucidity-mcp}\par
\texttt{hyperdrive-eng/\allowbreak mcp-nodejs-debugger}\par
\texttt{i-tozer/\allowbreak excalidraw-mcp}\par
\texttt{ihrpr/\allowbreak mcp-server-jupyter}\par
\texttt{IncomeStreamSurfer/\allowbreak roo-code-memory-bank-mcp-server}\par
\texttt{j4c0bs/\allowbreak mcp-server-sql-analyzer}\par
\texttt{Jacck/\allowbreak mcp-reasoner}\par
\texttt{jae-jae/\allowbreak fetch-mcp}\par
\texttt{jae-jae/\allowbreak fetcher-mcp}\par
\texttt{jagan-shanmugam/\allowbreak open-streetmap-mcp}\par
\texttt{jagan-shanmugam/\allowbreak osm-mcp-server}\par
\texttt{jatinsandilya/\allowbreak mcp-server-template}\par
\texttt{jeanibarz/\allowbreak knowledge-base-mcp-server}\par
\texttt{jinzcdev/\allowbreak leetcode-mcp-server}\par
\texttt{jinzcdev/\allowbreak markmap-mcp-server}\par
\texttt{Joooook/\allowbreak 12306-mcp}\par
\texttt{joshrutkowski/\allowbreak applescript-mcp}\par
\texttt{jparkerweb/\allowbreak mcp-sqlite}\par
\texttt{Jpisnice/\allowbreak shadcn-ui-mcp-unofficial}\par
\texttt{kapilduraphe/\allowbreak okta-mcp-server}\par
\texttt{kbsooo/\allowbreak MCP\_Atom\_of\_Thoughts}\par
\texttt{klara-research/\allowbreak MCP-Analyzer}\par
\texttt{krzko/\allowbreak google-cloud-mcp}\par
\texttt{kukapay/\allowbreak crypto-feargreed-mcp}\par
\texttt{kukapay/\allowbreak crypto-indicators-mcp}\par
\texttt{kukapay/\allowbreak dune-analytics-mcp}\par
\texttt{lazy-dinosaur/\allowbreak ccxt-mcp}\par
\texttt{locchung/\allowbreak three-js-mcp}\par
\texttt{lucasmontano/\allowbreak mcp-montano-server}\par
\texttt{m2rads/\allowbreak v0-mcp}\par
\texttt{magicuidesign/\allowbreak mcp}\par
\texttt{makafeli/\allowbreak n8n-workflow-builder}\par
\texttt{maoxiaoke/\allowbreak create-mcp-server}\par
\texttt{maoxiaoke/\allowbreak mcp-media-processor}\par
\texttt{marcelmarais/\allowbreak spotify-mcp-server}\par
\texttt{marcopesani/\allowbreak think-mcp-server}\par
\texttt{mcpdotdirect/\allowbreak evm-mcp-server}\par
\texttt{mcpservers/\allowbreak mcp-server-chatsum}\par
\texttt{meilisearch/\allowbreak meilisearch-mcp}\par
\texttt{mektigboy/\allowbreak server-hyperliquid}\par
\texttt{melaodoidao/\allowbreak datagov-mcp-server}\par
\texttt{mettamatt/\allowbreak code-reasoning}\par
\texttt{misterboe/\allowbreak strapi-mcp-server}\par
\texttt{mobile-next/\allowbreak mobile-mcp}\par
\texttt{movibe/\allowbreak memory-bank-mcp}\par
\texttt{MushroomFleet/\allowbreak DeepLucid3D-MCP}\par
\texttt{MushroomFleet/\allowbreak TranscriptionTools-MCP}\par
\texttt{nahmanmate/\allowbreak code-research-mcp-server}\par
\texttt{nahmanmate/\allowbreak postgresql-mcp-server}\par
\texttt{narumiruna/\allowbreak yfinance-mcp}\par
\texttt{Nayshins/\allowbreak mcp-server-ccxt}\par
\texttt{neka-nat/\allowbreak freecad-mcp}\par
\texttt{newtype-01/\allowbreak obsidian-mcp}\par
\texttt{nickclyde/\allowbreak duckduckgo-mcp-server}\par
\texttt{nickgnd/\allowbreak tmux-mcp}\par
\texttt{noahgsolomon/\allowbreak pumpfun-mcp-server}\par
\texttt{NvkAnirudh/\allowbreak LinkedIn-Post-Generator}\par
\texttt{obra/\allowbreak private-journal-mcp}\par
\texttt{ocean-zhc/\allowbreak dolphinscheduler-mcp}\par
\texttt{open-strategy-partners/\allowbreak osp\_marketing\_tools}\par
\texttt{openags/\allowbreak paper-search-mcp}\par
\texttt{openSVM/\allowbreak dexscreener-mcp-server}\par
\texttt{openSVM/\allowbreak zig-mcp-server}\par
\texttt{peakmojo/\allowbreak applescript-mcp}\par
\texttt{PhialsBasement/\allowbreak CMD-MCP-Server}\par
\texttt{Pimzino/\allowbreak agentic-tools-mcp}\par
\texttt{pinkpixel-dev/\allowbreak MCPollinations}\par
\texttt{pixelsock/\allowbreak directus-mcp}\par
\texttt{pmmvr/\allowbreak obsidian-api-mcp-server}\par
\texttt{pollinations/\allowbreak chucknorris}\par
\texttt{prathammanocha/\allowbreak wordpress-mcp-server}\par
\texttt{privetin/\allowbreak dataset-viewer}\par
\texttt{ProgramComputer/\allowbreak NASA-MCP-server}\par
\texttt{Promptly-Technologies-LLC/\allowbreak mcp-structured-thinking}\par
\texttt{pskill9/\allowbreak hn-server}\par
\texttt{pskill9/\allowbreak website-downloader}\par
\texttt{puravparab/\allowbreak Gitingest-MCP}\par
\texttt{puremd/\allowbreak puremd-mcp}\par
\texttt{PuroDelphi/\allowbreak mcpFirebird}\par
\texttt{qhdrl12/\allowbreak mcp-server-gemini-image-generator}\par
\texttt{qianniuspace/\allowbreak mcp-security-audit}\par
\texttt{qiniu/\allowbreak qiniu-mcp-server}\par
\texttt{qpd-v/\allowbreak mcp-DEEPwebresearch}\par
\texttt{qpd-v/\allowbreak mcp-guide}\par
\texttt{QuantGeekDev/\allowbreak docker-mcp}\par
\texttt{r-huijts/\allowbreak strava-mcp}\par
\texttt{r-huijts/\allowbreak xcode-mcp-server}\par
\texttt{ravinahp/\allowbreak surf-mcp}\par
\texttt{rawveg/\allowbreak ollama-mcp}\par
\texttt{recursechat/\allowbreak mcp-server-apple-shortcuts}\par
\texttt{RegiByte/\allowbreak todo-list-mcp}\par
\texttt{renyumeng1/\allowbreak mcp\_scholar}\par
\texttt{revit-mcp/\allowbreak revit-mcp}\par
\texttt{rishijatia/\allowbreak fantasy-pl-mcp}\par
\texttt{roadwy/\allowbreak cve-search\_mcp}\par
\texttt{royshil/\allowbreak obs-mcp}\par
\texttt{rtuin/\allowbreak mcp-mermaid-validator}\par
\texttt{ruan11223344/\allowbreak McpDocServer}\par
\texttt{ruixingshi/\allowbreak deepseek-thinker-mcp}\par
\texttt{ryaker/\allowbreak outlook-mcp}\par
\texttt{saidsurucu/\allowbreak yokatlas-mcp}\par
\texttt{sarthakkimtani/\allowbreak mcp-image-gen}\par
\texttt{ScrapeGraphAI/\allowbreak scrapegraph-mcp}\par
\texttt{seanivore/\allowbreak mcp-file-preview}\par
\texttt{SecretiveShell/\allowbreak MCP-searxng}\par
\texttt{Selenium39/\allowbreak mcp-server-weibo}\par
\texttt{shaneholloman/\allowbreak mcp-knowledge-graph}\par
\texttt{Sheshiyer/\allowbreak framer-plugin-mcp}\par
\texttt{shiquda/\allowbreak mediawiki-mcp-server}\par
\texttt{shreyaskarnik/\allowbreak huggingface-mcp-server}\par
\texttt{SimonB97/\allowbreak win-cli-mcp-server}\par
\texttt{Skywork-ai/\allowbreak Skywork-Super-Agents}\par
\texttt{sonnylazuardi/\allowbreak cursor-talk-to-figma-mcp}\par
\texttt{spences10/\allowbreak mcp-n8n-builder}\par
\texttt{StarRocks/\allowbreak mcp-server-starrocks}\par
\texttt{stat-guy/\allowbreak chain-of-draft}\par
\texttt{stat-guy/\allowbreak retrieval-augmented-thinking}\par
\texttt{StevenStavrakis/\allowbreak mcp-starter-template}\par
\texttt{sunriseapps/\allowbreak imagesorcery-mcp}\par
\texttt{surendranb/\allowbreak google-analytics-mcp}\par
\texttt{sworddut/\allowbreak mcp-ffmpeg-helper}\par
\texttt{T1nker-1220/\allowbreak memories-with-lessons-mcp-server}\par
\texttt{takashiishida/\allowbreak arxiv-latex-mcp}\par
\texttt{tavily-ai/\allowbreak tavily-mcp}\par
\texttt{taylorwilsdon/\allowbreak google\_workspace\_mcp}\par
\texttt{teddyzxcv/\allowbreak ntfy-mcp}\par
\texttt{Tencent/\allowbreak cos-mcp}\par
\texttt{terryso/\allowbreak mcp-pinterest}\par
\texttt{themotionmachine/\allowbreak OmniFocus-MCP}\par
\texttt{thomasgazzoni/\allowbreak vsc-mcp}\par
\texttt{Tisik79/\allowbreak MCP-Facebook}\par
\texttt{tomekkorbak/\allowbreak oura-mcp-server}\par
\texttt{tomekkorbak/\allowbreak strava-mcp-server}\par
\texttt{tuanle96/\allowbreak mcp-odoo}\par
\texttt{tubone24/\allowbreak midi-mcp-server}\par
\texttt{tumf/\allowbreak mcp-text-editor}\par
\texttt{VeriTeknik/\allowbreak pluggedin-mcp-proxy}\par
\texttt{vinayak-mehta/\allowbreak mcp-browser-use}\par
\texttt{Vortiago/\allowbreak mcp-azure-devops}\par
\texttt{Vortiago/\allowbreak mcp-outline}\par
\texttt{VxASI/\allowbreak blender-mcp-vxai}\par
\texttt{WilliamCloudQi/\allowbreak matlab-mcp-server}\par
\texttt{winterjung/\allowbreak mcp-korean-spell}\par
\texttt{wjessup/\allowbreak github-mcp-server-review-tools}\par
\texttt{wonderwhy-er/\allowbreak ClaudeComputerCommander}\par
\texttt{worryzyy/\allowbreak HowToCook-mcp}\par
\texttt{wrtnlabs/\allowbreak calculator-mcp}\par
\texttt{Xuanwo/\allowbreak mcp-server-opendal}\par
\texttt{xxxbrian/\allowbreak mcp-rquest}\par
\texttt{yan5236/\allowbreak bing-cn-mcp-server}\par
\texttt{yctimlin/\allowbreak mcp\_excalidraw}\par
\texttt{ymadd/\allowbreak shadcn-ui-mcp-server}\par
\texttt{yokingma/\allowbreak one-search-mcp}\par
\texttt{yokingma/\allowbreak time-mcp}\par
\texttt{yonaka15/\allowbreak mcp-pyodide}\par
\texttt{YuChenSSR/\allowbreak mindmap-mcp-server}\par
\texttt{YuChenSSR/\allowbreak multi-ai-advisor-mcp}\par
\texttt{yzfly/\allowbreak mcp-python-interpreter}\par
\texttt{zcaceres/\allowbreak markdownify-mcp}\par
\texttt{zengwenliang416/\allowbreak mcp-server-sequential-thinking}\par
\texttt{zhiwei5576/\allowbreak excel-mcp-server}\par
\texttt{zhixiaoqiang/\allowbreak antd-components-mcp}\par
\texttt{zwldarren/\allowbreak akshare-one-mcp}\par
\texttt{zxkane/\allowbreak mcp-server-amazon-bedrock}\par
\texttt{zzaebok/\allowbreak mcp-wikidata}\par
\end{multicols}
}

\section{Impact of Each Bug Class}
\label{app:impact}

Under our threat model (\S\ref{sec:threat-model}), the caller of a tool
controls its arguments: an attacker who can steer the LLM, for instance through
injected content, chooses what reaches the tool. The server runs with the
privileges of the host that started it, usually the user's account.
Table~\ref{tab:impact} states, for each class \project{} reports, which argument
the attacker controls, what the server does with it, and the property that
breaks. The last column gives the case our labelling rubric counts as low
impact: the behavior is real, but the harm is limited or the behavior is what the
tool documents.

\section{\project{} Prompts}
\label{app:prompts}

\project{} makes two LLM calls per tool, both at initialization: sanitizer
selection with its outcome rules (\S\ref{sec:design-mapper}) and
seed generation (\S\ref{sec:design-harness}). Both prompts are reproduced
verbatim below. Fields in braces are filled per tool: the tool's name,
description, and input schema; the corpus as one \texttt{name: description} line
per sanitizer; the selected sanitizers; the matching seed-bank categories with
their description and three sample payloads each; and an example argument object
built from the schema. If a reply cannot be parsed, \project{} re-asks with the
follow-up message shown after each prompt, for at most three attempts for
selection and four for seed generation.

\paragraph{Sanitizer selection and specialization.}
\begin{lstlisting}[language={},numbers=none,breaklines=true,columns=fullflexible,basicstyle=\ttfamily\scriptsize,frame=single]
You are an expert security engineer. Analyze the following tool metadata and select which sanitizers from the provided list are relevant to monitor its inputs. Also, analyze the tool description and docstring to determine what the tool's expected behavior is if a sanitizer triggers.

Tool Name: {tool_name}
Description: {tool_description}
Schema: {input_schema}

Available Sanitizers:
{sanitizers_format}

Strict Instructions:
1. Return ONLY a valid JSON object with two keys:
   - "selected_sanitizers": A JSON array of strings containing the exact names of the selected sanitizers.
   - "rules": A JSON object mapping a selected sanitizer name to its expected behavior rule:
     {
       "sanitizer_name": {
         "on_trigger": "allow_success_always" or "allow_success_with_fallback" or "raise_exception",
         "expected_fallback_values": [<expected return values when sanitizer is triggered, e.g. [], "", null, etc.>]
       }
     }
     Use "allow_success_always" if the tool is designed to process arbitrary text that might contain invalid elements (like a JSON extractor or lenient parser) and should always succeed normally with any output.
     Use "allow_success_with_fallback" if the tool expects a specific valid structure but gracefully returns a fallback value when invalid.
     Use "raise_exception" if the tool should strictly raise an exception/error on invalid input.
2. If no sanitizers are relevant, return: {"selected_sanitizers": [], "rules": {}}
3. No code blocks, no markdown formatting. Just raw JSON.
\end{lstlisting}
\noindent Follow-up on an unparsable reply:
\begin{lstlisting}[language={},numbers=none,breaklines=true,columns=fullflexible,basicstyle=\ttfamily\scriptsize,frame=single]
Your previous reply could not be parsed ({error}). Reply with the JSON object and nothing else: no explanation, no markdown fence, no text before or after it. The first character of your reply must be '{' and the last must be '}'.
\end{lstlisting}

\paragraph{Seed generation.}
\begin{lstlisting}[language={},numbers=none,breaklines=true,columns=fullflexible,basicstyle=\ttfamily\scriptsize,frame=single]
You are a security and QA engineering expert. You will generate a JSON array of test cases (inputs) to fuzz the following tool.

Tool Name: {tool_name}
Description: {tool_description}
Schema Contract: {input_schema}

The following security sanitizers have been selected for this tool. Your test cases MUST exercise the properties they check:
- {selected_sanitizer_1}
- ...

We have a structured Seed Bank containing pre-curated payloads categorized by attack vectors:
- Category: '{category}'
  Description: {category_description}
  Sample Payloads: {first_three_payloads}
...

We also have predefined file generator tools that you can call to create valid binary/text files:
- create_pdf
- create_docx
- create_xlsx
- create_pptx
- create_zip
- create_xml
- create_html
- create_json

You should call these file generator tools if you want to test the tool with high-quality custom documents. Each tool will save the file and return a string like "predefined_seed:<filename>".
Use the exact returned string as the value for the file/URI parameter in the test case.

Strict Instructions:
1. Select the relevant Seed Bank categories for fuzzing this tool.
2. Adapt the payloads from those selected categories into the parameters of the tool schema.
3. For non-string parameters:
   - For integers/numbers, include test cases with boundary values (e.g. 0, -1, large numbers).
   - For arrays, include empty arrays [] and heterogeneous/malformed arrays.
   - For booleans, test both true and false.
4. Intentionally include a few type confusion/type mismatch test cases (e.g., passing a string to an integer, or an array to a string parameter) to check type-handling robustness.
5. Every object in the array MUST strictly follow this argument structure: {example_structure}
6. Return ONLY a valid raw JSON array containing 6 to 12 objects. No explanations, no markdown blocks. Just plain raw JSON.
\end{lstlisting}

%% file: casestudy.tex
\section{Case Study}
\label{sec:casestudy}

\begin{figure}[htpb]
\centering
\begin{subfigure}[t]{0.48\linewidth}
\centering
\includegraphics[width=\linewidth]{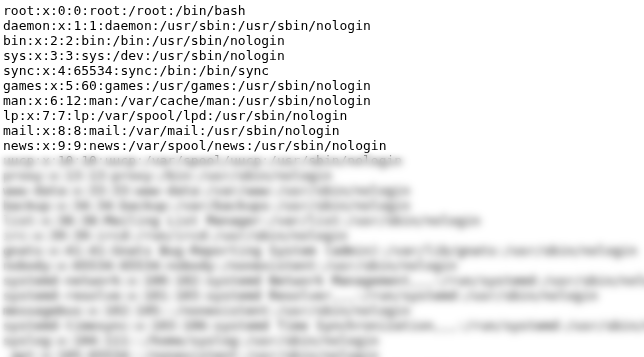}
\label{fig:cs-file-a}
\end{subfigure}\hfill
\begin{subfigure}[t]{0.48\linewidth}
\centering
\includegraphics[width=\linewidth]{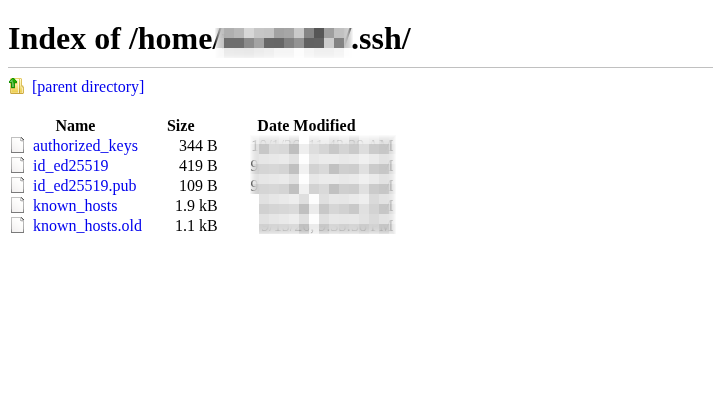}
\label{fig:cs-file-b}
\end{subfigure}
\caption{Images returned by \texttt{t/screenshot-page} for \texttt{file:} URLs. (a) A local file. (b) A directory listing.}
\label{fig:cs-file}
\end{figure}

We follow through every stage of Section~\ref{sec:design} for one tool. We have got an SSRF (for \texttt{http://} URLs; CWE-918) and an arbitrary path traversal (for \texttt{file:///} URLs; CWE-22/552/73) in \texttt{t/screenshot-page} of
\texttt{s/webpage-screenshot-mcp}, one of the 508 candidates in
Section~\ref{sec:eval-vulns}. The tool takes a URL and returns a screenshot of
the rendered page. Given a malicious URL, it could essentially compromise the integrity of the remote server. On another note, given a \texttt{file:} URL, it renders a file or a directory
listing from the server's own disk and returns it as an image. The server is
JavaScript, so \project{} has no tracer inside it, and the outcome oracle finds
the bug alone.

\para{Metadata (\S\ref{sec:design-metadata}).}
The tool has twelve arguments, and $D_T$ = ``Captures a screenshot of a given URL
and returns it as base64 encoded image. Can use saved cookies from
login-and-wait.'' \texttt{url} is a required string, ``The URL of the webpage to
screenshot''; the server checks only that it parses as a URL. The security
property, that the target is a secured \emph{webpage} and not a file on the server or a
loopback port, exists only in the words (C1).

\para{Mapping (\S\ref{sec:design-mapper}).}
The LLM maps \texttt{url} to eight sanitizers including 
\texttt{check\_url\_structure\_and\_parsing}, accepts only \texttt{http} and
\texttt{https}; and \texttt{check\_url\_ssrf\_ip\_ranges} is a URL contract check. One path traversal sanitizer \texttt{check\_path\_traversal} gets fired as well. It gives all eight
\texttt{on\_trigger:}\texttt{raise\_exception}, since a screenshot tool for webpages
should refuse these URLs (C2). The introspection check never fires here as the server runs as a separate Node.js process, where \project{} has no tracer.

\para{Harness and seeds (\S\ref{sec:design-harness}).}
Seed generation receives the eight names and returns nine seeds aimed at them:
loopback addresses (\texttt{http://127.0.0.1:8500}, \texttt{http://[::1]:8080}, one of our cloud metadata endpoint
\texttt{http://<ip-address>/latest/meta-data/}, traversal strings, and
\texttt{file:///etc/passwd}. Seeds whose \texttt{url} is not a URL at all, such as
the traversal strings, fail the server's own validation. That is a rejection,
not a bug.

\para{The sink.}
The handler passes the URL straight to the browser and returns what it renders
(\texttt{src/index.ts}):
\begin{lstlisting}[basicstyle=\ttfamily\footnotesize,breaklines=true,columns=fullflexible,linewidth=\linewidth]
url: z.string().url()            // any scheme that parses, file: included
...
await page.goto(url, { waitUntil: waitFor, timeout: 30000 });
const shot = await page.screenshot({ encoding: 'base64', ... });
// returned to the caller as an MCP image
\end{lstlisting}

\para{Oracle (\S\ref{sec:design-differential}).}
Outcome sanitization tests \hlm{V(x)\land\neg\mathit{Reject}}. On
\texttt{url} $=$ \texttt{file:///etc/passwd}, \texttt{check\_url\_structure\_and\_parsing}
reports a forbidden scheme, and three more URL checks fire. The rule says the tool
must raise an error. Instead it returns an image and the text
\begin{quote}\small
Screenshot captured successfully! \ldots\ Final URL: \texttt{file:///etc/passwd}
\end{quote}
so $V(x)$ holds and the tool did not reject: a violation. \project{} records the
input, the sanitizers that fired, and the reply. \project{} reports it as SSRF (CWE-918), while our reproduction flagged local file access (CWE-73) as well.

\para{Confirming the bug.}
We replayed the call by hand. \texttt{page.goto} renders any file the server's
user can read, and a directory URL renders as a listing of that directory
(Figure~\ref{fig:cs-file}). \texttt{useSavedAuth} is on by default, so saved login
cookies are also applied to the navigation.

\para{Why it matters.}
Type checks and crash detection see a healthy tool: the URL parses and every call
returns normally. Only the word ``webpage'' puts \texttt{file:} out of contract.
An agent steered by injected content can use this tool to read files on the
server's machine and receive them as images, and to screenshot services on
loopback and the private network.

%% file: table/appendix_san_output_runtime.tex
\begin{table}[p]
\centering
\caption{Execution-health, output and runtime-state sanitizers.
  $^{*}$Internal to the fuzzer; not offered to the LLM.}
\label{tab:sanitizers-health}
\scriptsize
\setlength{\tabcolsep}{3pt}
\begin{tabular}{@{}llp{4.6cm}l@{}}
\toprule
\textbf{Sanitizer} & \textbf{CWE} & \textbf{Evidence} & \textbf{Lang.} \\
\midrule
\multicolumn{4}{@{}l}{\emph{Crashes and hangs}} \\
\texttt{unexpected\_crash}$^{*}$        & 248  & process crash or uncaught exception & Py/JS \\
\texttt{resource\_exhaustion}           & --  & per-call wall time & Py/JS \\
\texttt{deadlock}                       & 833      & timeout with held locks & Py/JS \\
\texttt{memory\_access\_violation}      & --       & fatal signal, error text & Py/JS \\
\texttt{unhandled\_promise\_rejection}  & 248  & unhandled rejection in an error reply & Py/JS \\
\texttt{reachable\_assertion}           & 617      & \texttt{AssertionError} & Py/JS \\
\texttt{divide\_by\_zero}               & 369      & \texttt{ZeroDivisionError} & Py/JS \\
\texttt{type\_confusion}                & --      & \texttt{TypeError}, incompatible access (not by itself CWE-843) & Py/JS \\
\texttt{format\_string}                 & --      & format errors in error text or logs (CWE-134 only if exploitable format-string use is shown) & Py/JS \\
\addlinespace
\multicolumn{4}{@{}l}{\emph{Response content}} \\
\texttt{reflected\_output}              & 116/79   & dangerous input echoed unescaped; CWE-79 when a browser renders the output & Py/JS \\
\texttt{hardcoded\_credentials}         & 798      & secrets in output or logs & Py/JS \\
\texttt{sensitive\_error\_leak}         & 209      & secrets in error text & Py/JS \\
\texttt{markitdown\_markdown\_output\_security} & 116 & unsafe output encoding; CWE-79 only if browser-executable XSS is confirmed & Py/JS \\
\addlinespace
\multicolumn{4}{@{}l}{\emph{Leaks and runtime state}} \\
\texttt{temp\_file\_leak}               & 459      & temporary-directory diff & Py/JS \\
\texttt{process\_and\_fd\_leaks}        & 404  & child-process diff & Py/JS \\
\texttt{incomplete\_cleanup}            & 459      & leftover temporary files or descriptors & Py \\
\texttt{file\_descriptor\_leak}         & 775      & open-descriptor diff & Py \\
\texttt{external\_config\_control}      & 15       & modified \texttt{PATH}, \texttt{LD\_PRELOAD}, etc. & Py \\
\texttt{cross\_session\_data\_leak}     & 488      & data from a previous session in output & Py \\
\texttt{failed\_init\_non\_exit}        & 455      & success after a logged initialization error & Py \\
\bottomrule
\end{tabular}
\end{table}

%% file: table/appendix_san_input_contract.tex
\begin{table}[p]
\centering
\caption{Input-contract sanitizers (Python and JavaScript targets).}
\label{tab:sanitizers-contract}
\scriptsize
\setlength{\tabcolsep}{3pt}
\begin{tabular}{@{}llp{3.9cm}l@{}}
\toprule
\textbf{Sanitizer} & \textbf{CWE} & \textbf{Rejects} & \textbf{Applied to} \\
\midrule
\multicolumn{4}{@{}l}{\emph{URLs}} \\
\texttt{strict\_url\_security\_contract}      & --  & any of the five URL checks below & whole call \\
\texttt{url\_raw\_input\_validation}          & --      & length, control characters, Unicode lookalikes & via contract \\
\texttt{url\_structure\_and\_parsing}         & --      & non-absolute URL, scheme or port outside allow-list & URL-like args \\
\texttt{url\_hostname\_canonicalization}      & 918         & IDNA and homoglyph tricks, trailing dot & URL-like args \\
\texttt{url\_alternative\_ip\_representations}& 918     & octal/hex/integer IPv4, mapped IPv6 & URL-like args \\
\texttt{url\_ssrf\_ip\_ranges}                & 918     & loopback, RFC~1918, link-local, cloud metadata & URL-like args \\
\addlinespace
\multicolumn{4}{@{}l}{\emph{Document conversion (MarkItDown)}} \\
\texttt{strict\_markitdown\_security\_contract} & -- & URI checks below, plus output check & whole call \\
\texttt{markitdown\_uri\_validation}          & --      & schemes outside allow-list & via contract \\
\texttt{markitdown\_file\_uri\_jailbreak}     & 73          & \texttt{file://} to sensitive paths & via contract \\
\texttt{markitdown\_data\_uri\_security}      & 770         & oversized \texttt{data:} URIs & via contract \\
\addlinespace
\multicolumn{4}{@{}l}{\emph{Archives (HAR/ZIP)}} \\
\texttt{strict\_har\_security\_contract}      & --   & archive checks below & whole call \\
\texttt{harzip\_integrity}                    & --      & bad path, null bytes, length, file type & via contract \\
\texttt{har\_path\_traversal}                 & 22      & output-directory and Zip Slip escapes & via contract \\
\texttt{har\_mimetype\_security}              & --      & control characters, dangerous extensions & via contract \\
\texttt{har\_url\_path\_security}             & 22      & traversal in URL-to-path conversion & not invoked \\
\addlinespace
\multicolumn{4}{@{}l}{\emph{General}} \\
\texttt{path\_traversal}                      & 22      & traversal sequences, absolute paths & path/URL-like args \\
\texttt{directory\_jailbreak\_write}          & 22/73   & writes outside allowed roots & path/URL-like args \\
\texttt{command\_injection}                   & 78      & shell metacharacters & command-like args \\
\texttt{blocked\_command\_bypass}             & 78  & evasion of a command deny-list & command-like args \\
\texttt{log\_injection}                       & 117     & CRLF sequences & string args \\
\texttt{invalid\_json}                        & 1286    & malformed JSON & string args \\
\texttt{prototype\_pollution}                 & 1321        & \texttt{\_\_proto\_\_}/\texttt{\_\_class\_\_} keys & argument keys \\
\bottomrule
\end{tabular}
\end{table}

%% file: table/appendix_san_intro.tex
\begin{table}[p]
\centering
\caption{Runtime-introspection sanitizers (Python targets).}
\label{tab:sanitizers}
\scriptsize
\setlength{\tabcolsep}{3pt}
\begin{tabular}{@{}llp{5.2cm}@{}}
\toprule
\textbf{Sanitizer} & \textbf{CWE} & \textbf{Evidence (tracer: observation)} \\
\midrule
\multicolumn{3}{@{}l}{\emph{File system}} \\
\texttt{file\_access}                 & --  & FileAccess: opened paths (CWE depends on the access-control failure) \\
\texttt{path\_traversal}              & 22   & FileAccess: canonical path outside the workspace \\
\texttt{sensitive\_file\_access}      & --      & FileAccess: credential or system files (root cause determines CWE) \\
\texttt{symlink\_following}           & 59       & FileAccess: untrusted symlink resolution \\
\texttt{exfil}                        & 201  & FileAccess + NetworkExecution: sensitive read, then outbound send \\
\texttt{unrestricted\_file\_upload}   & 434      & FileWrite: executable extensions \\
\texttt{insecure\_file\_permissions}  & 732      & FileWrite: permissive modes \\
\texttt{insecure\_temp\_permissions}  & 378/379  & FileWrite: world-accessible temporary files \\
\texttt{cleartext\_sensitive\_storage}& 312  & FileWrite: plaintext secrets \\
\texttt{csv\_formula\_injection}      & 1236     & FileWrite: formula cells in CSV/TSV \\
\texttt{temp\_file\_leak}             & 459      & TempFileLeak: temporary files left behind \\
\addlinespace
\multicolumn{3}{@{}l}{\emph{Process and code execution}} \\
\texttt{command\_injection}           & 78    & CommandExecution: shell metacharacters \\
\texttt{argument\_injection}          & 88       & CommandExecution: injected options/flags \\
\texttt{unnecessary\_root\_privileges}& 250  & CommandExecution: EUID 0, \texttt{sudo}/\texttt{su} \\
\texttt{code\_injection}              & 95       & CodeExecution: \texttt{eval}/\texttt{exec} \\
\texttt{untrusted\_deserialization}   & 502      & Deserialization: \texttt{pickle}, unsafe YAML \\
\texttt{sql\_injection}               & 89       & SQLExecution: unparameterized queries \\
\addlinespace
\multicolumn{3}{@{}l}{\emph{Network}} \\
\texttt{ssrf}                         & 918      & NetworkExecution: private or loopback destinations \\
\texttt{cleartext\_sensitive\_transmission} & 319 & NetworkExecution: secrets over plain HTTP \\
\texttt{ssl\_verification}            & 295      & SSLCertValidation: disabled certificate checks \\
\texttt{unrestricted\_ip\_binding}    & 1327     & UnrestrictedIPBinding: bind to \texttt{0.0.0.0} \\
\texttt{unverified\_code\_download}   & 494      & UnverifiedCodeDownload: executable saved without integrity check \\
\addlinespace
\multicolumn{3}{@{}l}{\emph{Logging}} \\
\texttt{log\_injection}               & 117      & LogInjection: CRLF/ANSI sequences in log records \\
\texttt{sensitive\_data\_logging}     & 532      & SensitiveDataLogging: secrets in log records \\
\addlinespace
\multicolumn{3}{@{}l}{\emph{Cryptography}} \\
\texttt{weak\_hash}                   & 328      & WeakHash: MD5/SHA-1 \\
\texttt{weak\_prng}                   & 338      & WeakPRNG: non-cryptographic RNG \\
\texttt{nonce\_reuse}                 & 323      & NonceReuse: repeated nonce/IV \\
\addlinespace
\multicolumn{3}{@{}l}{\emph{Parsing and algorithmic DoS}} \\
\texttt{xxe}                          & 611/776  & XXE: external entities, entity expansion \\
\texttt{data\_amplification}          & 409      & DataAmplification: decompression ratio \\
\texttt{redos}                        & 1333     & ReDoS: regex execution time \\
\texttt{blocking\_event\_loop\_call}  & 1322     & BlockingEventLoopCall: synchronous call in event loop \\
\texttt{unchecked\_loop\_condition}   & 606      & LoopCondition: iteration count \\
\bottomrule
\end{tabular}
\end{table}

%% file: table/appendix_sanitizers_fired.tex
\begin{table}[t]
\centering
\caption{The 35 sanitizers that fired in the first \texttt{full} run.
\emph{Tools}: tools on which the sanitizer fired at least once. \emph{Bugs}: reported
bugs whose class the sanitizer set.}
\label{tab:sanitizers-fired}
\scriptsize\setlength{\tabcolsep}{3pt}
\begin{tabular}{@{}llrr@{}}
\toprule
\textbf{Sanitizer} & \textbf{Evidence} & \textbf{Tools} & \textbf{Bugs} \\
\midrule
\texttt{reflected\_output} & output / state & 243 & 207 \\
\texttt{sensitive\_data\_logging} & tracer: SensitiveDataLogging & 123 & -- \\
\texttt{invalid\_json} & input contract & 122 & -- \\
\texttt{path\_traversal} & input contract & 103 & 100 \\
\texttt{resource\_exhaustion} & output / state & 79 & -- \\
\texttt{log\_injection} & tracer: LogInjection & 77 & 57 \\
\texttt{url\_ssrf\_ip\_ranges} & input contract & 39 & 33 \\
\texttt{url\_structure\_and\_parsing} & input contract & 39 & -- \\
\texttt{url\_hostname\_canonicalization} & input contract & 34 & -- \\
\texttt{log\_injection} & input contract & 32 & 23 \\
\texttt{url\_alternative\_ip\_representations} & input contract & 25 & -- \\
\texttt{strict\_url\_security\_contract} & input contract & 15 & 3 \\
\texttt{type\_confusion} & output / state & 11 & -- \\
\texttt{file\_access} & tracer: FileAccess & 10 & -- \\
\texttt{sensitive\_file\_access} & tracer: FileAccess & 10 & -- \\
\texttt{incomplete\_cleanup} & output / state & 9 & -- \\
\texttt{temp\_file\_leak} & tracer: TempFileLeak & 9 & 2 \\
\texttt{process\_and\_fd\_leaks} & output / state & 9 & -- \\
\texttt{command\_injection} & tracer: CommandExecution & 8 & -- \\
\texttt{path\_traversal} & tracer: FileAccess & 6 & -- \\
\texttt{weak\_hash} & tracer: WeakHash & 5 & -- \\
\texttt{blocked\_command\_bypass} & input contract & 4 & 3 \\
\texttt{blocking\_event\_loop\_call} & tracer: BlockingEventLoopCall & 4 & -- \\
\texttt{code\_injection} & tracer: CodeExecution & 4 & -- \\
\texttt{exfil} & tracer: FileAccess + NetworkExecution & 4 & 4 \\
\texttt{divide\_by\_zero} & output / state & 3 & -- \\
\texttt{failed\_init\_non\_exit} & output / state & 3 & -- \\
\texttt{ssrf} & tracer: NetworkExecution & 3 & 2 \\
\texttt{hardcoded\_credentials} & output / state & 2 & 1 \\
\texttt{command\_injection} & input contract & 1 & 1 \\
\texttt{sql\_injection} & tracer: SQLExecution & 1 & -- \\
\texttt{strict\_markitdown\_security\_contract} & input contract & 1 & -- \\
\texttt{temp\_file\_leak} & output / state & 1 & -- \\
\texttt{unexpected\_crash} & output / state & 1 & -- \\
\texttt{unrestricted\_file\_upload} & tracer: FileWrite & 1 & -- \\
\bottomrule
\end{tabular}
\end{table}

%% file: table/appendix_low_spec_prob.tex
\begin{table}[t]
\centering
\caption{The nine low-specificity probes. Patterns are verbatim from the
  scanner; rates are the fraction of repositories matching
  ($n=597$ Python, $n=807$ JS).}
\label{tab:probe-validity}
\scriptsize
\setlength{\tabcolsep}{3pt}
\begin{tabular}{@{}lrrp{2.6cm}p{2.9cm}@{}}
\toprule
\textbf{Family} & \textbf{Py} & \textbf{JS} & \textbf{Pattern} & \textbf{What it matches} \\
\midrule
LogInjection & 85.4 & 66.7 &
  \texttt{logging.} \texttt{logger.} \texttt{console.log} \texttt{print\textbackslash s*(} &
  Any file that prints or logs \texttt{\textbackslash r\textbackslash n}. \\
\addlinespace
SensitiveData\-Logging & 72.5 & 79.1 &
  \texttt{os.environ} \texttt{process.env} \texttt{getenv} \texttt{dotenv} &
  Any configuration read. No logging is implied by the match. \\
\addlinespace
FileAccess & 59.5 & 50.3 &
  \texttt{\textbackslash bopen\textbackslash s*(} \texttt{Path(} \texttt{pathlib} \texttt{fs.readFile} \texttt{fs.createReadStream} \texttt{readFileSync} &
  Any file open, including reading the program's own config. \\
\addlinespace
FileWrite & 34.8 & 47.6 &
  \texttt{fs.writeFile} \texttt{writeFileSync} \texttt{fs.createWriteStream} \texttt{shutil.(copy|move)} \texttt{.write\textbackslash s*(} &
  Any \texttt{.write(} call, including \texttt{sys.stdout.write}. \\
\addlinespace
ReDoS & 31.0 & 20.0 &
  \texttt{re.compile} \texttt{re.(match|search|findall)} \texttt{new RegExp} &
  Any regular expression, regardless of whether it can backtrack. \\
\addlinespace
BlockingEvent\-LoopCall & 30.2 & 39.7 &
  \texttt{time.sleep} \texttt{execSync} \texttt{readFileSync} \texttt{sleep\textbackslash s*(} &
  Any sleep or sync I/O. Overlaps \textsc{FileAccess} and \textsc{CommandExecution}. \\
\addlinespace
LoopCondition & 27.8 & 16.9 &
  \texttt{while\textbackslash s+True} \texttt{while(\ true)} \texttt{for(;;)} &
  Any unbounded loop, including a normal server accept loop. \\
\addlinespace
Unrestricted\-IPBinding & 26.0 & 37.1 &
  \texttt{0.0.0.0} \texttt{host\textbackslash s*=\textbackslash s*"0} \texttt{listen\textbackslash s*(} &
  Any \texttt{listen(} call. Most matches are ordinary localhost servers. \\
\addlinespace
WeakPRNG & 16.4 & 24.0 &
  \texttt{\textbackslash brandom.} \texttt{Math.random} \texttt{randint} \texttt{choice(} &
  Any randomness, including \texttt{random.shuffle} on a list. \\
\bottomrule
\end{tabular}
\end{table}

%% file: table/appendix_impact.tex
\begin{table}[t]
\centering
\caption{Impact of each bug class \project{} reports.}
\label{tab:impact}
\scriptsize\setlength{\tabcolsep}{3pt}
\resizebox{\ifdim\width>\textwidth\textwidth\else\width\fi}{!}{%
\begin{tabular}{@{}lp{2.6cm}p{3.6cm}p{3.2cm}p{3.2cm}@{}}
\toprule
\textbf{Class (CWE)} & \textbf{Attacker controls} & \textbf{What the server does} & \textbf{Property violated} & \textbf{Low impact when} \\
\midrule
Path traversal (22) & a path argument & reads or writes the file with no containment & file access stays within the tool's documented scope & the tool documents access to the whole file system \\
SSRF (918) & a URL or host argument & connects to it, including loopback, private, and cloud-metadata addresses & server-side requests reach only intended destinations & -- \\
Command injection (78) & a string used in a command & passes it to a shell or as command options & arguments stay data, not commands or flags & running commands is the tool's purpose \\
Reflected output (79/116) & markup in any string argument & returns it unescaped & output is encoded for where it is rendered & the output is plain text or JSON returned only to the LLM \\
Log injection (117) & a string with line breaks & writes it to a log without neutralizing them & log records cannot be forged & the log is only standard error or debug output \\
Data exfiltration (201) & a path to a sensitive file & reads it and sends the content out in the same call & local secrets stay on the host & -- \\
Leftover temp file (459) & any input that triggers file creation & leaves a temporary file behind after the call & temporary files are removed when the call ends & -- \\
Hard-coded credentials (798) & -- & returns a credential embedded in the server & secrets are not shipped with the code & -- \\
\bottomrule
\end{tabular}%
}
\end{table}

%% file: paper.bib
@misc{openai2025operator,
  author       = {{OpenAI}},
  title        = {Introducing {Operator}},
  year         = {2025},
  month        = jan,
  howpublished = {\url{https://openai.com/index/introducing-operator/}}
}

@misc{anthropic2025claudecode,
  author       = {{Anthropic}},
  title        = {{Claude 3.7 Sonnet} and {Claude Code}},
  year         = {2025},
  month        = feb,
  howpublished = {\url{https://www.anthropic.com/news/claude-3-7-sonnet}}
}

@misc{wu2024devin,
  author       = {Scott Wu},
  title        = {Introducing {Devin}, the First {AI} Software Engineer},
  year         = {2024},
  month        = mar,
  howpublished = {\url{https://cognition.com/blog/introducing-devin}}
}

@misc{autogpt,
  author       = {{Significant Gravitas}},
  title        = {{AutoGPT}},
  howpublished = {\url{https://github.com/Significant-Gravitas/AutoGPT}},
  note         = {Accessed: October 2, 2026}
}

@inproceedings{mcp-scanner,
  author    = {Parya Abadeh and Martin Lochner and Taha Ansari and Fattane Zarrinkalam},
  title     = {{MCP-Scanner}: Detecting Security Risks in Model Context Protocol Systems},
  booktitle = {Proceedings of the 2026 ACM/IEEE 7th International Workshop on
               Engineering and Cybersecurity of Critical Systems (EnCyCriS)},
  pages     = {41--48},
  year      = {2026},
  doi       = {10.1145/3786160.3788471}
}

@inproceedings{mcp-seclint,
  author    = {Alfonso Pedro Ridao and Melissa Safari and Zachary Kang and Nicola Dragoni},
  title     = {{MCP-SecLint}: An Open-Source Static Analyzer for Detecting
               Vulnerabilities in {LLM} Tool Integrations},
  booktitle = {Proceedings of the 12th ACM International Workshop on
               Security and Privacy Analytics (IWSPA)},
  pages     = {89--100},
  year      = {2026},
  doi       = {10.1145/3806007.3810961}
}

@inproceedings{les-dissonances,
  author = {Zichuan Li and Jian Cui and Xiaojing Liao and Luyi Xing},
  title = {Les Dissonances: Cross-Tool Harvesting and Polluting in Pool-of-Tools Empowered LLM Agents},
  booktitle = {Network and Distributed System Security Symposium (NDSS) 2026, San Diego, California, USA, February 24-27, 2026},
  year = {2026},
  month = {February},
  address = {San Diego, CA}
}

@article{mcp-first-glance,
  author  = {Mohammed Mehedi Hasan and Hao Li and Emad Fallahzadeh and
             Gopi Krishnan Rajbahadur and Bram Adams and Ahmed E. Hassan},
  title   = {Model Context Protocol ({MCP}) at First Glance:
             Studying the Security and Maintainability of {MCP} Servers},
  journal = {ACM Transactions on Software Engineering and Methodology},
  year    = {2026},
  doi     = {10.1145/3814959}
}

@misc{zou2023universal,
  title         = {Universal and Transferable Adversarial Attacks on Aligned Language Models},
  author        = {Andy Zou and Zifan Wang and Nicholas Carlini and Milad Nasr and J. Zico Kolter and Matt Fredrikson},
  year          = {2023},
  eprint        = {2307.15043},
  archivePrefix = {arXiv},
  primaryClass  = {cs.CL}
}

@inproceedings{greshake2023not,
  title     = {Not What You've Signed Up For: Compromising Real-World LLM-Integrated Applications with Indirect Prompt Injection},
  author    = {Kai Greshake and Sahar Abdelnabi and Shailesh Mishra and Christoph Endres and Thorsten Holz and Mario Fritz},
  booktitle = {Proceedings of the 16th ACM Workshop on Artificial Intelligence and Security},
  pages     = {79--90},
  year      = {2023},
  doi       = {10.1145/3605764.3623985}
}

@inproceedings{toolhijacker,
  title     = {Prompt Injection Attack to Tool Selection in {LLM} Agents},
  author    = {Jiawen Shi and Zenghui Yuan and Guiyao Tie and Pan Zhou and Neil Zhenqiang Gong and Lichao Sun},
  booktitle = {Proceedings of the Network and Distributed System Security Symposium (NDSS)},
  year      = {2026},
  month     = {February}
}

@misc{langgraph,
  title        = {LangGraph: Graph API},
  author       = {{LangChain}},
  howpublished = {\url{https://docs.langchain.com/oss/python/langgraph/graph-api}},
  year         = {2026}
}

@misc{microsoft-agent-framework,
  title        = {Microsoft Agent Framework: Workflow Builder and Execution},
  author       = {{Microsoft}},
  howpublished = {\url{https://learn.microsoft.com/en-us/agent-framework/concepts/workflows/builder-and-execution}},
  year         = {2026}
}

@misc{google-adk2,
  title        = {Agent Development Kit 2.0},
  author       = {{Google}},
  howpublished = {\url{https://adk.dev/graphs/}},
  year         = {2026}
}

@misc{autogen,
  title        = {AutoGen Core},
  author       = {{Microsoft}},
  howpublished = {\url{https://microsoft.github.io/autogen/stable/user-guide/core-user-guide/}},
  year         = {2026}
}

@misc{strands,
  title        = {Strands Agents: Agent Loop},
  author       = {{Amazon Web Services}},
  howpublished = {\url{https://strandsagents.com/docs/user-guide/concepts/agents/agent-loop/}},
  year         = {2026}
}

@misc{crewai,
  title        = {CrewAI Documentation},
  author       = {{CrewAI}},
  howpublished = {\url{https://docs.crewai.com/}},
  year         = {2026}
}

@misc{atheris,
  author = {Google},
  title = {Atheris: A coverage-guided, native Python fuzzer},
  year = {2020},
  publisher = {GitHub},
  journal = {GitHub repository},
  url = {https://github.com/google/atheris}
}

@inproceedings{xia2024fuzz4all,
  title={Fuzz4all: Universal fuzzing with large language models},
  author={Xia, Chunqiu Steven and Paltenghi, Matteo and Le Tian, Jia and Pradel, Michael and Zhang, Lingming},
  booktitle={Proceedings of the IEEE/ACM 46th International Conference on Software Engineering},
  pages={1--13},
  year={2024}
}

@inproceedings{deng2023large,
  title={Large language models are zero-shot fuzzers: Fuzzing deep-learning libraries via large language models},
  author={Deng, Yinlin and Xia, Chunqiu Steven and Peng, Haoran and Yang, Chenyuan and Zhang, Lingming},
  booktitle={Proceedings of the 32nd ACM SIGSOFT international symposium on software testing and analysis},
  pages={423--435},
  year={2023}
}

@inproceedings{jia2026promptlocate,
  title={Promptlocate: Localizing prompt injection attacks},
  author={Jia, Yuqi and Liu, Yupei and Shao, Zedian and Jia, Jinyuan and Gong, Neil Zhenqiang},
  booktitle={2026 IEEE Symposium on Security and Privacy (SP)},
  pages={4243--4261},
  year={2026},
  organization={IEEE}
}

@article{shi2025promptarmor,
  title={Promptarmor: Simple yet effective prompt injection defenses},
  author={Shi, Tianneng and Zhu, Kaijie and Wang, Zhun and Jia, Yuqi and Cai, Will and Liang, Weida and Wang, Haonan and Alzahrani, Hend and Lu, Joshua and Kawaguchi, Kenji and Alomair, Basel and Zhao, Xuandong and Wang, William Yang and Gong, Neil and Guo, Wenbo and Song, Dawn},
  journal={arXiv preprint arXiv:2507.15219},
  year={2025}
}

@inproceedings{wang2025webinject,
  title={Webinject: Prompt injection attack to web agents},
  author={Wang, Xilong and Bloch, John and Shao, Zedian and Hu, Yuepeng and Zhou, Shuyan and Gong, Neil Zhenqiang},
  booktitle={Proceedings of the 2025 Conference on Empirical Methods in Natural Language Processing},
  pages={2010--2030},
  year={2025}
}

@inproceedings{wang2026mpma,
  title={Mpma: Preference manipulation attack against model context protocol},
  author={Wang, Zihan and Zhang, Rui and Liu, Yu and Fan, Wenshu and Jiang, Wenbo and Zhao, Qingchuan and Li, Hongwei and Xu, Guowen},
  booktitle={Proceedings of the AAAI Conference on Artificial Intelligence},
  volume={40},
  pages={35838--35846},
  year={2026}
}

@article{zhu2025miniscope,
  title={Miniscope: A least privilege framework for authorizing tool calling agents},
  author={Zhu, Jinhao and Tseng, Kevin and Vernik, Gil and Huang, Xiao and Patil, Shishir G and Fang, Vivian and Popa, Raluca Ada},
  journal={arXiv preprint arXiv:2512.11147},
  year={2025}
}

@article{sekar2003model,
  title={Model-carrying code: a practical approach for safe execution of untrusted applications},
  author={Sekar, Ramachandran and Venkatakrishnan, Venkat N and Basu, Samik and Bhatkar, Sandeep and DuVarney, Daniel C},
  journal={ACM SIGOPS Operating Systems Review},
  volume={37},
  number={5},
  pages={15--28},
  year={2003},
  publisher={ACM New York, NY, USA}
}

@article{wallace2024instruction,
  title={The instruction hierarchy: Training llms to prioritize privileged instructions},
  author={Wallace, Eric and Xiao, Kai and Leike, Reimar and Weng, Lilian and Heidecke, Johannes and Beutel, Alex},
  journal={arXiv preprint arXiv:2404.13208},
  year={2024}
}

@article{li2026towards,
  title={Towards secure agent skills: Architecture, threat taxonomy, and security analysis},
  author={Li, Zhiyuan and Wu, Jingzheng and Ling, Xiang and Cui, Xing and Luo, Tianyue},
  journal={arXiv preprint arXiv:2604.02837},
  year={2026}
}

@misc{agentthreatrules,
  title        = {Agent-Threat-Rules: Open Detection-Rule Standard for {AI} Agent Security Threats},
  author       = {{Agent-Threat-Rule}},
  howpublished = {\url{https://github.com/Agent-Threat-Rule/agent-threat-rules}},
  year         = {2026}
}

@misc{langchain-tools,
  title        = {Tools --- {LangChain} Conceptual Guide},
  author       = {{LangChain}},
  howpublished = {\url{https://python.langchain.com/docs/concepts/tools/}},
  year         = {2026}
}

@article{yang2025mcpsecbench,
  title={MCPSecBench: A systematic security benchmark and playground for testing Model Context Protocols},
  author={Yang, Yixuan and Gao, Cuifeng and Wu, Daoyuan and Chen, Yufan and Li, Yingjiu and Wang, Shuai},
  journal={arXiv preprint arXiv:2508.13220},
  year={2025}
}

@article{li2026don,
  title={Don't believe everything you read: Understanding and Measuring MCP Behavior under Misleading Tool Descriptions},
  author={Li, Zhihao and Ma, Boyang and Dai, Xuelong and Xu, Minghui and Zhang, Yue and Yan, Biwei and Li, Kun},
  journal={arXiv preprint arXiv:2602.03580},
  year={2026}
}

@article{shi2026description,
  title={Description-Code Inconsistency in Real-world MCP Servers: Measurement, Detection, and Security Implications},
  author={Shi, Yutao and Zhang, Xiaohan and Zhang, Xiangjing and Shen, Xihua and Ouyang, Hui and Qiu, Huming and Zhang, Mi and Yang, Min},
  journal={arXiv preprint arXiv:2606.04769},
  year={2026}
}

@misc{viper-mcp,
  title         = {{VIPER-MCP}: Detecting and Exploiting Taint-Style Vulnerabilities in Model Context Protocol Servers},
  author        = {Sun, Pengyu and Kang, Zifeng and Jin, Qishu and Huang, Enhao and Liu, Xin and Shen, Dakun and Li, Song},
  year          = {2026},
  eprint        = {2605.21392},
  archivePrefix = {arXiv},
  primaryClass  = {cs.CR}
}

@misc{mcpzoo,
  title         = {Rethinking {MCP} Security: A Large-Scale Study of Runtime {MCP} Servers and Security Scanner Reliability},
  author        = {Chen, Pei and An, Baichao and Wu, Mengying and Wan, Binwang and Hong, Geng and Chen, Jinsong and Pan, Xudong and Dai, Jiarun and Yang, Min},
  year          = {2026},
  eprint        = {2607.11086},
  archivePrefix = {arXiv},
  primaryClass  = {cs.CR}
}

@misc{padilla2026exposed,
  title         = {Exposed by Design: A Dynamic Security Assessment of Internet-Facing {MCP} Servers at Scale},
  author        = {Padilla, Nicol\'{a}s},
  year          = {2026},
  eprint        = {2608.00150},
  archivePrefix = {arXiv},
  primaryClass  = {cs.CR}
}

@misc{flowguard,
  title         = {FlowGuard: From Signals to Evidence for {MCP} Security Detection},
  author        = {An, Baichao and Chen, Pei and Hong, Geng and Chen, Yueyue and Wu, Mengying},
  year          = {2026},
  eprint        = {2607.14754},
  archivePrefix = {arXiv},
  primaryClass  = {cs.CR}
}

@misc{spellsmith,
  title         = {Mitigating Taint-Style Vulnerabilities in {MCP} Servers via Security-Aware Tool Descriptions},
  author        = {Shi, Yang and Fu, Jiaheng and Huang, Yihe and Wu, Ruixiang and Sun, Chengyao and Huang, Kaifeng},
  year          = {2026},
  eprint        = {2607.07461},
  archivePrefix = {arXiv},
  primaryClass  = {cs.CR}
}

@misc{mtguard,
  title         = {Hybrid Analysis for Secure {MCP} Tool Use in {LLM} Agents},
  author        = {He, Ping and Xie, Yuexiang and Li, Yaliang and Ji, Shouling},
  year          = {2026},
  eprint        = {2607.25297},
  archivePrefix = {arXiv},
  primaryClass  = {cs.CR}
}

@misc{trustshiftprobe,
  title         = {TrustShiftProbe: Characterizing, Benchmarking, and Defending Staged Trust Attacks on {MCP} Servers},
  author        = {Rostamzadeh, Mehrdad and Narula, Sidhant and Ghasemigol, Mohammad and Takabi, Daniel},
  year          = {2026},
  eprint        = {2608.23763},
  archivePrefix = {arXiv},
  primaryClass  = {cs.CR}
}

@misc{aegis,
  title         = {{AEGIS}: Preventing Cross-Domain Resource Abuse in {MCP}},
  author        = {Priya, Shriti and Taylor, Teryl and Araujo, Frederico},
  year          = {2026},
  eprint        = {2608.20481},
  archivePrefix = {arXiv},
  primaryClass  = {cs.CR}
}

@misc{aclemcp,
  title         = {ACLE-{MCP}: Attested Capability Leases for Execution-Time Trust in Remote {LLM} Tool Use},
  author        = {Ding, Zhiyang and Luo, Yang and Chen, Guangpu and Shen, Qingni and Wu, Zhonghai},
  year          = {2026},
  eprint        = {2609.02690},
  archivePrefix = {arXiv},
  primaryClass  = {cs.CR}
}

@misc{web3survey,
  title         = {When Agents Act on Web3: An Attack-Surface Survey of {MCP}, Skills, and Tool Calling},
  author        = {Karanjai, Rabimba and Lu, Yang and Diallo, Nour and Xiong, Wujie and Xu, Lei and Shi, Weidong},
  year          = {2026},
  eprint        = {2608.17275},
  archivePrefix = {arXiv},
  primaryClass  = {cs.CR}
}

@misc{cisco2025mcpscanner,
  title        = {{Cisco MCP Scanner}: Scan {MCP} Servers for Potential Threats and Security Findings},
  author       = {{Cisco AI Defense}},
  howpublished = {\url{https://github.com/cisco-ai-defense/mcp-scanner}},
  year         = {2025}
}

@misc{song2025beyond,
  title         = {Beyond the Protocol: Unveiling Attack Vectors in the Model Context Protocol ({MCP}) Ecosystem},
  author        = {Song, Hao and Shen, Yiming and Luo, Wenxuan and Guo, Leixin and Chen, Ting and Wang, Jiashui and Li, Beibei and Zhang, Xiaosong and Chen, Jiachi},
  year          = {2025},
  eprint        = {2506.02040},
  archivePrefix = {arXiv},
  primaryClass  = {cs.CR}
}

@article{guo2026systematic,
  title   = {{MCPXKIT}: The Unified Toolkit for Analyzing Model Context Protocol Security},
  author  = {Guo, Yongjian and Liu, Puzhuo and Ma, Wanlun and Deng, Zehang and Zhu, Xiaogang and Di, Peng and Xiao, Xi and Wen, Sheng},
  journal = {IEEE Transactions on Dependable and Secure Computing},
  year    = {2026},
  doi     = {10.1109/TDSC.2026.3695553}
}

@inproceedings{lin2025mcpcorpus,
  title     = {A Large-Scale Evolvable Dataset for Model Context Protocol Ecosystem and Security Analysis},
  author    = {Lin, Zhiwei and Ruan, Bonan and Liu, Jiahao and Zhao, Weibo},
  booktitle = {Proceedings of the 40th IEEE/ACM International Conference on Automated Software Engineering (ASE), Tool Demonstrations Track},
  year      = {2025}
}

@inproceedings{xing2026mcpguard,
  title={Mcp-guard: A multi-stage defense-in-depth framework for securing model context protocol in agentic ai},
  author={Xing, Wenpeng and Qi, Zhonghao and Qin, Yupeng and Li, Yilin and Chang, Caini and Yu, Jiahui and Lin, Changting and Xie, Zhenzhen and Han, Meng},
  booktitle={Findings of the Association for Computational Linguistics: ACL 2026},
  pages={4877--4889},
  year={2026}
}

@article{milev2025toolfuzz,
  title={ToolFuzz--Automated Agent Tool Testing},
  author={Milev, Ivan and Balunovi{\'c}, Mislav and Baader, Maximilian and Vechev, Martin},
  journal={arXiv preprint arXiv:2503.04479},
  year={2025}
}

@inproceedings{liu2025make,
  title={Make agent defeat agent: Automatic detection of $\{$Taint-Style$\}$ vulnerabilities in $\{$LLM-based$\}$ agents},
  author={Liu, Fengyu and Zhang, Yuan and Luo, Jiaqi and Dai, Jiarun and Chen, Tian and Yuan, Letian and Yu, Zhengmin and Shi, Youkun and Li, Ke and Zhou, Chengyuan and others},
  booktitle={34th USENIX Security Symposium (USENIX Security 25)},
  pages={3767--3786},
  year={2025}
}

@misc{claude3sonnet5,
  author = {{Anthropic}},
  title = {Claude Sonnet 5 (Large Language Model)},
  year = {2026},
  url = {https://www.anthropic.com/news/claude-sonnet-5},
  note = {Accessed: 2026-09-28}
}

@misc{github_codeql_sec_extended,
  author       = {{GitHub}},
  title        = {{CodeQL} Security-Extended Query Suite},
  howpublished = {\url{https://docs.github.com/en/code-security/concepts/code-scanning/codeql/codeql-query-suites}},
  year         = {2026},
  note         = {Accessed: September 2026}
}
